%% file: tesis_root.tex
\documentclass[letterpaper,12pt]{report}
\usepackage{amsfonts} % para incluir fonts y
\usepackage{amssymb}  % simbolos de la AMS
\usepackage{graphicx} % standard LaTeX graphics tool
\usepackage{amsmath}  % paquete matemático
\usepackage[spanish]{babel}
\usepackage[latin1]{inputenc}
\usepackage{latexsym}
\usepackage[title]{appendix}
\usepackage[Algoritmo]{algorithm}
\usepackage{algorithmic}
\newtheorem{theorem}{Teorema}[chapter]

\newcommand{\EndProof}{\hfill $\blacksquare$}

\renewcommand{\bibname}{Bibliograf\'{\i}a}

\newcommand{\R}{{\cal {R}}}
\newcommand{\HH}{{\cal {H}}}
\begin{document}
\pagestyle{empty}
\begin{titlepage}
\begin{center}
{\large \bf UNIVERSIDAD DE LA HABANA}

\vspace{0.3cm}

{\large \bf Facultad de F\'isica}

\vspace{0.5cm}

\vspace{2in}

{\Huge \bf Estabilidad de \'atomos en el dominio ani\'onico ($Z<N$) }\\

\vspace{1in}

{\large \bf Tesis de Diploma}\\

\vspace{0.8cm}

%{\large  \it presentada en opci\'on al grado de} \\
%{\Large  \bf Master en Ciencias F\'{\i}sicas}\\

\vspace{1.2in}

{\large \bf Autor:}\hspace{0.3cm}{\large Gabriel Jos\'e Gil P\'erez} \\

\vspace{0.2in}

{\large \bf Tutor:} \hspace{0.3cm}{\large Dr.Cs. Augusto de Jes\'us Gonz\'alez Garc\'ia, ICIMAF.}

\vspace{2in}

{\large 22 de mayo de 2011, La Habana, Cuba}

\end{center}

\end{titlepage}

%%%%%%%%%%%%%%%%%%%%%%%%%%%%%%%%%%%%%%%%%%%%%%%%%%%%%%%%%%%%%%%
%	Exergo

\newpage
\voffset=2in %\thispagestyle{empty} \hoffset=3in

\begin{flushright}

{\Huge \it \ \ \ \small{(...) La base de su aritm\'etica es la noci\'on de n\'umeros indefinidos. (...) Afirman que la operaci\'on de contar modifica las cantidades y las convierte de indefinidas en definidas. El hecho de que varios individuos que cuentan una misma cantidad logran un resultado igual, es para los psic\'ologos un ejemplo de asociaci\'on de ideas o de buen ejercicio de la memoria. Ya sabemos que en Tl\"{o}n el sujeto del conocimiento es uno y eterno.}}

\vspace{0.1in}

{\Large \it \small{Tl\"{o}n, Uqbar, Orbis Tertius, Jorge Luis Borges}}

\vspace{0.5in}

{\Huge \it \ \ \ \small{If in some cataclysm, all scientific knowledge were to be destroyed, and only one sentence passed on to the next generations of creatures, what statement would contain the most information in the fewest words? I believe it is atomic hypothesis (or atomic fact, or whatever you wish to call it) that all things are made of atoms...}}\\

\vspace{0.1in}

{\Large \it \small{The Feynman Lectures on Physics, Chapter 1: ``Atoms in motion", Richard P. Feynman  }}

\end{flushright}

%%%%%%%%%%%%%%%%%%%%%%%%%%%%%%%%%%%%%%%%%%%%%%%%%%%%%%%%%%%%%%%
%	Dedicatoria

\newpage
\voffset=2in %\thispagestyle{empty} \hoffset=3in

\begin{flushright}

{\Huge \it \ \ \ \small{A mi padres,}}\\

{\Large \it \small{por su eterno sacrificio, su amor y su luz...}}\\

\vspace{0.2in}

{\Huge \it \ \ \ \small{A Laura,}}\\

{\Large \it \small{a quien le debo el coraz\'on central, que no comercia con palabras...}}

\end{flushright}

\newpage
\voffset=0in % Margen superior de las páginas 1in por default+voffset

\pagestyle{headings} \pagenumbering{roman} \setcounter{page}{1}

%%%%%%%%%%%%%%%%%%%%%%%%%%%%%%%%%%%%%%%%%%%%%%%%%%%%%%%%%%%%%%%%%%%
% Agradecimientos

\input{agradecimientos.tex}

%%%%%%%%%%%%%%%%%%%%%%%%%%%%%%%%%%%%%%%%%%%%%%%%%%%%%%%%%%%%%%%%%%%
% Resumen

\input{resumen.tex}

%%%%%%%%%%%%%%%%%%%%%%%%%%%%%%%%%%%%%%%%%%%%%%%%%%%%%%%%%%%%%%%%
\newpage
\voffset=0in % Margen superior de las páginas 1in por default+voffset

\pagestyle{headings} %\pagenumbering{roman} \setcounter{page}{1}

\tableofcontents \pagebreak \pagenumbering{arabic}
\setcounter{page}{1}

%%%%%%%%%%%%%%%%%%%%%%%%%%%%%%%%%%%%%%%%%%%%%%%%%%%%%%%%%%%%%%%%%%%
%Introducción
\input{intro.tex}

%%%%%%%%%%%%%%%%%%%%%%%%%%%%%%%%%%%%%%%%%%%%%%%%%%%%%%%%%%%%%%%%%%%
% Capitulo 1
%\input{Cap_1.tex}
\input{preliminares.tex}

%%%%%%%%%%%%%%%%%%%%%%%%%%%%%%%%%%%%%%%%%%%%%%%%%%%%%%%%%%%%%%%%%%%
% Capitulo 2
\input{cap_2.tex}

%%%%%%%%%%%%%%%%%%%%%%%%%%%%%%%%%%%%%%%%%%%%%%%%%%%%%%%%%%%%%%%%%%%
% Capitulo 3
\input{cap_3.tex}

%%%%%%%%%%%%%%%%%%%%%%%%%%%%%%%%%%%%%%%%%%%%%%%%%%%%%%%%%%%%%%%%%%%
% Conclusiones
\input{conclusiones.tex}

%%%%%%%%%%%%%%%%%%%%%%%%%%%%%%%%%%%%%%%%%%%%%%%%%%%%%%%%%%%%%%%%%%%
% Recomendaciones
\input{recomendaciones.tex}

%%%%%%%%%%%%%%%%%%%%%%%%%%%%%%%%%%%%%%%%%%%%%%%%%%%%%%%%%%%%%%%%%%%
% Anexos

\begin{appendices}
 \noappendicestocpagenum
 \addappheadtotoc
 \input{appendixa.tex}
 \input{appendixc.tex}
 \input{appendixb.tex}
\end{appendices}

%%%%%%%%%%%%%%%%%%%%%%%%%%%%%%%%%%%%%%%%%%%%%%%%%%%%%%%%%%%%%%%%%%%
% Referencias

\input{bib.tex}
%%%%%%%%%%%%%%%%%%%%%%%%%%%%%%%%%%%%%%%%%%%%%%%%%%%%%%%%%%%%%%%%%%%

\end{document}

%% file: agradecimientos.tex
\chapter*{Agradecimientos}
%\addcontentsline{toc}{chapter}{Agradecimientos}
\markright{AGRADECIMIENTOS}

Estas, son siempre unas p\'aginas inc\'omodamente gratas, por oximor\'onica que resulte la frase. Uno debe hacer un gesto de reciprocidad que no debe parecer insuficiente y debe ser exhaustivo -nombrar a todos los que deben ser nombrados- y excluyente -nombrar s\'olo a aquellos que deben ser nombrados. Tal es una proeza que no me siento capaz de cumplir con acierto, pero como de remedos est\'a hecha nuestra existencia... aqu\'i est\'a:

A mi mam\'a, Aurora P\'erez, que siempre ha sabido el camino, por no cesar en el intento de indicarlo. A mi pap\'a, Jorge Gil, cuyos consejos y apoyo resultan inapreciables, por ense\~narme a ser observador y paciente. A ambos, mi ejemplo, que han forjado para m\'i el presente y el futuro y me aligeran el trayecto cada d\'ia.

A mis abuelos, Hugo P\'erez, Aurora Mart\'inez y Raquel Mateos, por su participaci\'on indispensable en mi educaci\'on, por su cari\~no y su entereza.

A mi t\'io Oscar P\'erez, al que hubiera querido demostrar mi admiraci\'on...

A mi primo Celso Hugo P\'erez, por ayudarnos mutuamente y por hacerme mejor persona en el proceso.

A mi t\'ia Olguita y a mi primo Oscar\'in, por estar siempre pendiente de los progresos de esta tesis.

A Laura Azor, por ser mi consejera en todas las materias, por su inteligencia emocional, su constancia y su determinaci\'on, pero, sobre todo, por hacer de m\'i un hombre feliz y pleno.

A mis suegros, Jorge Azor y Mayrena Hern\'andez, por soportar a un aprendiz de f\'isico en la familia, y por considerarme parte de ella.

A mis amigos del Taller Literario de Ciencia Ficci\'on Espacio Abierto, por tantas alegr\'ias, y por entender que el tiempo es relativo al observador...

A mis amigos de la Lenin, Michel, Alejandro, y Anabel, por su cercan\'ia a\'un cuando medien leguas de distancia.

Alguien dijo una vez que nadie quiere deberle a sus contempor\'aneos; yo estoy completamente en desacuerdo: no puedo hacer menos que expresar mi gratitud infinita a mi grupo de graduaci\'on y a los que nos acompa\~naron en el trayecto. Ante todo a Maikel y Manuel, por tanto estudio, por tantas experiencias juntos; pero tambi\'en a Ismael, L\'idice, Ferm\'in, Lupe, Yanela, Patricia, Rosa, Carlos Iv\'an, un grupo hasta el final de las consecuencias.

A mis profesores de la Facultad de Matem\'atica, Mario Armando G\'omez y Jos\'e Talavera, por su compromiso con la profesi\'on y por intentar inculc\'arnoslo.

A todos los profesores de la Facultad de F\'isica, en especial a Esperanza Pur\'on, Miguel Ramos, Mar\'ia S\'anchez Colina, Manuel Hern\'andez Calvi\~no, Soto, Julio C\'esar Drake, Roberto Mulet, Jos\'e Mar\'in, Melqu\'iades de Dios, y Mar\'ia Teresa P\'erez, que me ense\~naron m\'as que f\'isica y nos hicieron sentirnos parte de la Facultad, pese a la destrucci\'on/reconstrucci\'on del edificio. A los profesores j\'ovenes, Maurice Oliva, Alejandro Lage y Yuriel N\'u\~nez, que completaron nuestra formaci\'on y nos indujeron a entrenar la intuici\'on y la creatividad en la actividad cient\'ifica.

A mi tutor, Augusto Gonz\'alez, al que tengo como paradigma no s\'olo por su enorme ta\-len\-to y experiencia, sino por su voluntad, su perseverancia, y su obstinado \'impetu, cualidades que a mi juicio son inherentes a un cient\'ifico exitoso. A \'el le debo la conducci\'on de esta tesis por los derroteros correctos: desde la implementaci\'on de las herramientas computacionales hasta los c\'alculos n\'umericos espec\'ificos y su interpretaci\'on; desde la escritura del documento hasta el entrenamiento en exposiciones.

A Alain Delgado, por ayudarme con la programaci\'on en FORTRAN y con el sistema operativo Linux, por darme una explicaci\'on exhaustiva del m\'etodo de Interacci\'on de Configuraciones y su implementaci\'on particular en puntos cu\'anticos, y por garantizar parte del hardware necesario para completar los c\'alculos num\'ericos costos\'isimos de la tesis. A Alexander Odriazola, por su disposici\'on a ayudarme con FORTRAN, con Linux, con LaTex y con Grace, por sugerirme referencias bibliogr\'aficas, y por sus valiosos consejos.

A David, por colaborar en la puesta a punto de las m\'aquinas para calcular y por las innumerables consultas sobre Linux. 

Al Departamento de F\'isica Te\'orica, en particular, y al ICIMAF, en general, donde he respirado un ambiente de trabajo y camarader\'ia que trasciende los marcos de la ciencia.

%% file: resumen.tex
\chapter*{S\'intesis}
%\addcontentsline{toc}{chapter}{S\'intesis}
\markright{S\'INTESIS}

Se estudia el comportamiento de la Energ\'ia de
Ionizaci\'on ($E_{I}$) de \'atomos aislados con $N$ electrones como funci\'on de la
carga nuclear $Z$, considerada un par\'ametro arbitrario, en la regi\'on
$Z<N$. Se analizan \'atomos con la estructura electr\'onica de gases
nobles ($N=2, 10, 18$) y de metales alcalinos ($N=3, 11, 19$)
utilizando m\'etodos num\'ericos (Interacci\'on de Configuraciones con
excitaciones dobles) y modelos semi-anal\'iticos.

%% file: intro.tex
%%%%%%%%%%%%%%%%%%%%%%%%%%%%%%%%%%%%%%%%%%%%%%
%              Introducción                  %
%%%%%%%%%%%%%%%%%%%%%%%%%%%%%%%%%%%%%%%%%%%%%%

\chapter*{Introducción}
\addcontentsline{toc}{chapter}{Introducci\'on}
\markright{INTRODUCCI\'ON}

Desde principios del siglo XX, cuando se acumularon, como evidencia indirecta de la hip\'otesis at\'omica, el \'exito de la Teor\'ia Cin\'etica Molecular, la Tabla Peri\'odica de Mendel\'eev, el descubrimiento de iones y electrones, y la comprensi\'on del origen de la radiaci\'on electromagn\'etica, se comenzaron a estudiar, de manera te\'orica y experimental, los \'atomos como constituyentes fundamentales de la materia. Los trabajos de Thomson (1897), Rutherford (1911), Bohr (1913) y muchos otros intentaron describir la estructura interna del \'atomo y su interacci\'on con la radiaci\'on.

No obstante, el impulso m\'as significativo de la f\'isica at\'omica est\'a dado por el surgimiento de la Mec\'anica Cu\'antica, cuyas figuras m\'as importantes fueron de Broglie, Heisenberg, Schr\"{o}dinger, Born, Pauli y Dirac. Con esta teor\'ia, que prove\'ia una f\'isica conceptualmente diferente y un arsenal de nuevas herramientas te\'oricas, se pudieron responder muchas de las interrogantes planteadas sobre los fen\'omenos at\'omicos y se concibieron interpretaciones m\'as apropiadas sobre el tema; en particular, se explic\'o la naturaleza de los enlaces qu\'imicos y la no intuitiva estabilidad de aniones at\'omicos (iones at\'omicos con carga negativa). Uno de los resultados m\'as satisfactorios de la Mec\'anica Cu\'antica es la explicaci\'on l\'ogica del sistema peri\'odico de los elementos.

A partir de la d\'ecada del 30 se desarrollaron poderosos m\'etodos e instrumentos experimentales sobre la base de ideas cu\'anticas b\'asicas, entre ellos la resonancia magn\'etica de haces moleculares, el m\'aser y el l\'aser. Las nuevas t\'ecnicas fueron aplicadas a problemas b\'asicos de la electrodin\'amica cu\'antica, para el estudio de propiedades at\'omicas o nucleares y para el dise\~no de dispositivos como los relojes at\'omicos. 

Algunos avances subsecuentes de la f\'isica at\'omica est\'an relacionados con el control del movimiento y los estados internos de \'atomos e iones; el m\'etodo de espectroscop\'ia de radiofrecuencia con \'atomos excitados \'opticamente (1953), la t\'ecnica de trampa de iones (en la d\'ecada del 50), el enfriamiento Dopler (1978), y, m\'as recientemente, la condensaci\'on de Bose-Einstein en un gas at\'omico (1995) y la medici\'on del corrimiento al rojo gravitacional por interferometr\'ia at\'omica (2010), se cuentan entre los m\'as prominentes. Otros t\'opicos actuales de intensa labor investigativa son los estados transientes y la din\'amica de colisiones, la teor\'ia relativista de muchos cuerpos, la correlaci\'on electr\'onica, los iones m\'ultiplemente cargados y el estudio de \'atomos sometidos a intensos campos magn\'eticos.

Un resumen de la actividad cient\'ifica del siglo XX enmarcada en la f\'isica at\'omica se puede encontrar en el art\'iculo de Kleppner (\cite{0}). 

Dentro del estudio de los \'atomos, desde hace alg\'un tiempo, existe un inter\'es marcado en la estabilidad de aniones at\'omicos en trabajos tanto te\'oricos (\cite{1}-\cite{15}) como experimentales (\cite{11}, \cite{exp_1}, \cite{exp_2}).

Las investigaciones experimentales sobre el tema se concentran, en su mayor\'ia, en la b\'usqueda de aniones at\'omicos en fase gaseosa de un largo tiempo de vida, con $t \sim 1 \mu s$ (\cite{exp_2}), aunque tambi\'en existen trabajos que describen la estructura electr\'onica de dichos sistemas (\cite{exp_1}).

En el caso de estudios te\'oricos se suelen emplear modelos semi-anal\'iticos de una part\'icula (\cite{1}, \cite{9}-\cite{11}), m\'etodos \textit{ab initio} (\cite{2}, \cite{3}) caracter\'isticos de la Qu\'imica Cu\'antica, la teor\'ia de Lieb (\cite{1}, \cite{magnetic_1}, \cite{4}) basada en aplicar el principio variacional a la matriz densidad, descripciones an\'alogas a las transiciones de fase y fen\'omenos cr\'iticos (\cite{5}-\cite{11}) usuales en la F\'isica Estad\'istica, o aparatos matem\'aticos m\'as rigurosos (\cite{13}-\cite{15}) en el marco de la Mec\'anica Cu\'antica no relativista.

Estos trabajos generalmente no estudian la din\'amica del problema sino procesos estacionarios, de manera que el criterio de estabilidad no viene dado por el tiempo de vida de las especies en cuesti\'on sino por el hecho de que el estado del electr\'on m\'as lejano al n\'ucleo deje de ser ligado. Para este caso se tiene que la energ\'ia del sistema de $N-1$ electrones $E(Z,N-1)$ debe ser menor o igual que la energ\'ia del sistema con $N$ electrones $E(Z,N)$, o lo que es lo mismo la energ\'ia de ionizaci\'on del sistema debe ser nula o positiva. 

Fenomenol\'ogicamente, se puede entender que cuando se a\~naden electrones al \'atomo neutro el apantallamiento de carga hace que el n\'ucleo efectivo que ``percibe'' el \'ultimo electr\'on tenga carga nuclear cero o incluso menor, form\'andose una barrera de Coulomb repulsiva. Cuando esta barrera es suficientemente grande, y la repulsi\'on coulombiana es mayor que otras fuerzas atractivas debidas a los efectos de llenado de capas, el nivel de energ\'ia del electr\'on m\'as lejano se aproxima al \textit{continuum} y este se escapa por efecto t\'unel a trav\'es de la barrera. En estas condiciones se dice que el \'atomo se ``autoioniza'', i.e. pierde un electr\'on sin que se le suministre energ\'ia al sistema, o sea que la configuraci\'on del \'atomo con $N$ electrones resulta inestable debido a que existe un configuraci\'on con menor n\'umero de electrones que tiene energ\'ia igual o menor.

Existe una carga nuclear cr\'itica $Z^{(c)}$ para la cual ocurre este proceso de autoionizaci\'on y este valor es tal que se cumple que $E(Z^{(c)},N)=E(Z^{(c)},N-1)$. La comparaci\'on de $Z^{(c)}$ con el n\'umero de electrones $N$ del sistema puede ser usada en el criterio de estabilidad de aniones m\'ultiplemente cargados, como veremos m\'as adelante.

%Existe una carga nuclear cr\'itica $Z^{(c)}$ para la cual ocurre este proceso de autoionizaci\'on y este valor es tal que se cumple que $E(Z^{(c)},N)=E(Z^{(c)},N-1)$. Esta carga nuclear cr\'itica puede ser usada en el criterio de estabilidad de aniones at\'omicos m\'ultiplemente cargados: si $Z^{(c)} > N-1$ no existir\'an aniones at\'omicos, la energ\'ia de un \'atomo neutro es menor a la energ\'ia del sistema simplemente cargado; si $Z^{(c)} < N-1$ existe de manera estable el sistema con $N-1$ electrones, o sea la energ\'ia de un \'atomo neutro es mayor a la energ\'ia del sistema simplemente cargado; y si $Z^{(c)} < N-2$ existe el ani\'on doblemente cargado estable, la energ\'ia de un \'atomo neutro es mayor a la energ\'ia del sistema doblemente cargado.

Algunos art\'iculos (\cite{8}, \cite{11}) reportan valores de la carga nuclear
cr\'itica $Z^{(c)}$ para \'atomos con $N$ electrones para $2<N<86$, y demuestran la estabilidad de aniones at\'omicos simples para ciertos valores de $N$. Adem\'as, a pesar de que no exista un resultado de una teor\'ia de primeros principios que lo prohiba, existe consenso en la comunidad cient\'ifica relacionada con el tema acerca de la no existencia de aniones at\'omicos doblemente cargados (AAD) estables en fase gaseosa. C\'alculos \textit{ab initio} han comprobado muchas veces esta hip\'otesis (\cite{11}). Es conocido tambi\'en que los AAD pueden estabilizarse bajo la influencia de campos externos intensos (\cite{1},\cite{magnetic_1}, \cite{9}-\cite{12}, \cite{magnetic_2}), en fase gaseosa, o con la presencia de un medio solvente como el agua (\cite{2}). Para mol\'eculas s\'i se han obtenido evidencias de aniones doblemente cargados incluso de forma experimental (\cite{17}), y la b\'usqueda de aniones m\'ultiplemente cargados en mol\'eculas peque\~nas contin\'ua siendo de inter\'es en esta rama de la f\'isica (\cite{17},\cite{small_mol}).

Este trabajo tiene como motivaciones todos los estudios citados, en especial los trabajos de Kais y/o Serra (\cite{5}-\cite{12}) sobre la analog\'ia entre las transiciones de fase de la F\'isica Estad\'istica y el comportamiento del \'ultimo electr\'on en un \'atomo con carga nuclear cercana a $Z^{(c)}$. Estos autores demostraron que, para \'atomos con $N=2$, la energ\'ia de ionizaci\'on ($E_{I}$) en funci\'on de la carga efectiva del n\'ucleo ($Z_{core}=Z-N+1$), i.e. la carga del cori\'on compuesto por el n\'ucleo y los electrones internos, alrededor de ${Z_{core}}^{(c)}=Z^{(c)}-N+1$, exhibe una discontinuidad en su derivada, mientras que, para \'atomos con $N=3$, la aproximaci\'on de $E_{I}$ a cero es suave (\cite{11}). En la teor\'ia de transiciones de fase se dir\'ia que la transici\'on de fase en funci\'on del par\'ametro $Z_{core}$, an\'alogo de la temperatura, en el caso con $N=2$ es de primer orden mientras que en el de $N=3$ es de segundo orden o continua (\cite{11}). Una extensi\'on de este resultado podr\'ia ser, como se ha propuesto la presente tesis, probar que estas diferencias existen en general entre los \'atomos con la estructura electr\'onica de los metales alcalinos (Grupo IA de la Tabla Peri\'odica) y los que tienen capas electr\'onicas llenas, con la estructura electr\'onica de gases nobles (Grupo VIIIA de la Tabla Peri\'odica).

Por otro lado, para el dominio cati\'onico ($Z>N$ \'o $Z_{core}>1$) de iones at\'omicos se ha demostrado, utilizando la teor\'ia de Thomas-Fermi, la universalidad exis\-ten\-te en el comportamiento de la energ\'ia de ionizaci\'on (\cite{18}). Ser\'ia interesante, adem\'as, demostrar un resultado similar para la dominio ani\'onico e interpolar entre estas soluciones para obtener una curva universal de energ\'ia de ionizaci\'on de \'atomos para todo el rango de $Z$.

En este trabajo nos concentramos en el estudio de  la energ\'ia de ionizaci\'on de \'atomos e iones at\'omicos con $N=2, 10, 18$ (con la estructura electr\'onica de gases nobles) y $N=3, 11, 19$ (con la estructura electr\'onica de metales alcalinos), variando $Z_{core}$ en el rango $(-1,1)$, es decir en el dominio ani\'onico, con dos objetivos fundamentales: 

\begin{itemize}

\item Buscar relaciones universales en el comportamiento de la energ\'ia de ionizaci\'on cerca del valor umbral de la carga nuclear $Z^{(c)}$

\item Analizar la estabilidad de aniones at\'omicos en fase gasesosa en funci\'on del n\'umero de electrones del sistema.

\end{itemize}

Los c\'alculos num\'ericos necesarios se realizaron mediante el m\'etodo de Interacci\'on de Configuraciones hasta exci\-ta\-cio\-nes dobles (2p2h) utilizando como referencia la funci\'on de onda de Hartree-Fock. Los detalles de estos m\'etodos son descritos en el Cap\'itulo \ref{metodos} de manera exhaustiva, y las particularidades de los c\'alculos de la energ\'ia del estado b\'asico de \'atomos pueden encontrarse en el ep\'igrafe \ref{comp}.

Para describir la variaci\'on de $E_{I}$ con $Z_{core}$ se propuso un potencial simple (compuesto del pozo finito con profundidad $V_{0}$ y radio $r_{0}$ y el potencial de Coulomb con carga efectiva $Z_{core}$) que reproduce la interacci\'on del electr\'on d\'ebilmente ligado con el sistema a corto y a largo alcance. Modelos similares de una part\'icula se han planteado en otros trabajos para describir el \'atomo de $N$ electrones (\cite{1},\cite{9}-\cite{11}). En el ep\'igrafe \ref{modelo} se abunda sobre este modelo de una part\'icula y las soluciones de la Ecuaci\'on de Schr\"{o}dinger para el potencial planteado.

El trabajo presentado como tesis de diploma consta de una introducci\'on, tres capítulos, y  las correspondientes conclusiones. El contenido está distribuido de la siguiente forma:

\begin{itemize}

\item{\textbf{Introducci\'on:}} Se muestra una rese\~na hist\'orica de los avances fundamentales de la f\'isica at\'omica y se presenta el tema de estudio: la estabilidad de aniones at\'omicos en fase gaseosa.

\item{\textbf{Capítulo 1:}} Se discute el criterio de estabilidad empleado en funci\'on de la carga nuclear cr\'itica, se muestran caracter\'isticas generales de los metales alcalinos y los gases nobles, y se propone un m\'etodo semi-anal\'itico de una part\'icula que puede describir la energ\'ia de ionizaci\'on de estos sistemas electr\'onicos.

\item{\textbf{Capítulo 2:}} Se describen las caracter\'isticas fundamentales de los m\'etodos empleados para el c\'alculo num\'erico de la energ\'ia del estado b\'asico de \'atomos con $N$ electrones. 

\item{\textbf{Capítulo 3:}} Se se\~nalan los detalles computacionales del c\'alculo de la energ\'ia del estado b\'asico de \'atomos y se muestran los resultados obtenidos para sistemas at\'omicos con la estructura electr\'onica de los gases nobles y de los metales alcalinos. Se analiza la estabilidad de aniones at\'omicos y la universalidad en las curvas de energ\'ia de ionizaci\'on. Se obtiene una expresi\'on anal\'itica de la energ\'ia de ionizaci\'on del electr\'on m\'as lejano en funci\'on de la carga nuclear del cori\'on, empleando el modelo de una part\'icula propuesto.

\item{\textbf{Conclusiones:}} Se resaltan los resultados fundamentales y sus interpretaciones, haciendo \'enfasis en la estabilidad de los aniones y la universalidad de la energ\'ia de ionizaci\'on en el dominio ani\'onico para los \'atomos estudiados.

\item{\textbf{Recomendaciones:}} Se indican las posibles mejoras y extensiones que puede sufrir este trabajo en pos de convertirse en un estudio m\'as riguroso y exhaustivo. 

\end{itemize}

%% file: preliminares.tex
\chapter{Preliminares}
\label{preliminares}

\section{Energ\'ia de ionizaci\'on y estabilidad de \'atomos}
\label{criterio}

Dado que la energ\'ia de ionizaci\'on es una propiedad que se emplear\'a muy frecuentemente a lo largo de este trabajo, es preciso ofrecer una definici\'on rigurosa de ella:

\textbf{Definici\'on:} El potencial de ionizaci\'on ($I$) de un \'atomo es la energ\'ia que debe suministrarse al sistema para extraer de este un electr\'on, o, lo que es lo mismo, la energ\'ia necesaria para promover a un electr\'on desde el nivel de energ\'ia del espectro discreto en que se encuentra hasta el continuum. Se define, adem\'as, la energ\'ia de ionizaci\'on o de enlace ($E_{I}$), como la energ\'ia necesaria para que el \'atomo mantenga ligado al \'ultimo electr\'on ($E_{I}=-I$).
  
Esta energ\'ia se puede calcular para el caso de un \'atomo de $N$ electrones y carga nuclear $Z$ mediante la expresi\'on $E_{I}=E(Z,N)-E(Z,N-1)$, donde $E(Z,N)$ y $E(Z,N-1)$ son las energ\'ias del estado b\'asico del sistema con $N$ y $N-1$ part\'iculas, respectivamente. Como se afirm\'o en la introducci\'on de este trabajo, el criterio empleado para la estabilidad de un sistema at\'omico se refiere a la energ\'ia de ionizaci\'on del mismo, de manera que si el sistema tiene energ\'ia de ionizaci\'on $E_{I}<0$ es estable, mientras que si $E_{I}>0$ es inestable; la frontera entre estas regiones, $E_{I} \approx 0$ (umbral de autoionizaci\'on), ser\'a de particular importancia, en cap\'itulos posteriores (Ep. \ref{stability}), para el an\'alisis de universalidad.

La forma natural de entender la aproximaci\'on del sistema al umbral de autoionizaci\'on surge de analizar como se comportar\'ia la energ\'ia de ionizaci\'on al agregar electrones al \'atomo neutro. Es l\'ogico pensar en este caso que existir\'a un n\'umero m\'aximo de electrones que el n\'ucleo at\'omico pueda mantener ligados y, por tanto, toda configuraci\'on con un n\'umero mayor de electrones ser\'a inestable. Sin embargo, este no ser\'a el enfoque considerado en este trabajo, debido a que no es efectivo desde el punto de vista computacional\footnote{El valor del n\'umero de electrones $N$ aparece, en todas las expresiones de los m\'etodos de Qu\'imica Cu\'antica (v\'ease Cap\'itulo \ref{metodos}), como un l\'imite superior de las sumatorias (e.g. $\sum_{i,j:i<j}^{N}\frac{1}{|r_{i}-r{j}|}$), de manera que, a los efectos del c\'alculo num\'erico, carece de sentido extender el par\'ametro $N$ a valores no enteros. Por el contrario, para el par\'ametro $Z$ no existe esta dificultad (e.g. $-Z\sum_{i=1}^{N}\frac{1}{r_{i}}$), y, en teor\'ia, se podr\'ia calcular la energ\'ia de ionizaci\'on de un sistema con carga nuclear fraccionaria.}. Aqu\'i se tomar\'a como punto de partida el \'atomo neutro, como en el caso anterior, pero en vez de agregar electrones, se disminuir\'a de manera discreta (con paso $\Delta{Z}=0.1$) la carga nuclear $Z$, tomada como par\'ametro arbitrario, hasta que esta sea tal ($Z=Z^{(c)}$) que el n\'ucleo at\'omico sea incapaz de mantener ligado al \'ultimo electr\'on. (Se define la carga nuclear cr\'itica $Z^{(c)}$ como la carga nuclear tal que $E_{I}(Z^{(c)},N)=0$.). Aunque calcular la energ\'ia de sistemas con carga nuclear fraccionaria no tiene sentido f\'isico, por la cuantizaci\'on de la carga, este tratamiento resulta \'util para analizar estabilidad y universalidad de \'atomos, y ya ha sido empleado en los trabajos ($\cite{11}$) y ($\cite{13}$) con prop\'ositos similares.

De alguna manera resulta intuitivo que, a medida que el campo central del n\'ucleo se hace menos atractivo, por la disminuci\'on de la carga nuclear, el sistema se aproxima al umbral de autoionizaci\'on. O sea debe existir una carga nuclear cr\'itica $Z^{(c)}$ para la cual el sistema transita del estado con $N$ part\'iculas al estado con $N-1$ part\'iculas, tras el proceso de ionizaci\'on. Ahora bien, una demostraci\'on estricta de la existencia de una carga nuclear cr\'itica para un \'atomo de $N$ electrones es un poco m\'as elaborada y no ser\'a presentada en este trabajo, aunque puede encontrarse en el art\'iculo de Lieb (\cite{4}).

La comparaci\'on de la carga nuclear con el n\'umero de electrones del sistema analizado ofrece un criterio preciso sobre la estabilidad de un sistema. Para un \'atomo con $N$ electrones y carga $Q=Z-N$:

\begin{itemize}
\item si $Z^{(c)} \geq N-1$ no existe un ion negativo con carga $Q=-1$,
\item si $Z^{(c)} < N-1$ existe un ion negativo con carga $Q=-1$, y,
\item si $Z^{(c)} < N-2$ existe un ion negativo con carga $Q=-2$.
\end{itemize}

N\'otese que $Z^{(c)} < N$ debido a que el \'atomo neutro y los cationes at\'omicos ($Z \ge N$) son siempre estables (al menos para los valores de $Z$ que permitan la estabilidad del n\'ucleo at\'omico) su energ\'ia de ionizaci\'on es estrictamente positiva por tanto es necesario suministrarle energ\'ia al sistema para que sea ionizado.

Para analizar el comportamiento de la energ\'ia de ionizaci\'on, se requiere, como se apunt\'o anteriormente, la energ\'ia del estado b\'asico de los sistemas con $N$ y $N-1$ electrones. El c\'alculo de la energ\'ia y la funci\'on de onda del estado b\'asico, y de los estados excitados, es el objetivo de los m\'etodos de la Qu\'imica Cu\'antica, algunos de los cuales se explicar\'an m\'as adelante, en el Cap\'itulo \ref{metodos}. En este estudio se calcularon num\'ericamente estas magnitudes($E(Z,N)$, $E(Z,N-1)$) mediante el m\'etodo de Interacci\'on de Configuraciones hasta excitaciones dobles, al cual se dedica un ep\'igrafe del cap\'itulo siguiente (Ep. \ref{CI}).

\section{Metales alcalinos y gases nobles}
\label{alkalii_inert}

En un gr\'afico de $E_{I}$ con el n\'umero at\'omico $Z$ (con $Z=N$) los valores m\'as elevados resultan para los elementos de capas electr\'onicas llenas o gases nobles ($Z=2, 10, 18, 36, 54$, $86$), debido a la simetr\'ia esf\'erica de su estructura electr\'onica, mientras que, para los elementos que se encuentran inmediatamente despu\'es, denominados metales alcalinos ($Z=3, 11, 19, 37, 55, 87$), la ligadura del \'ultimo electr\'on es peque\~na y su energ\'ia de ionizaci\'on muy baja. En la energ\'ia de excitaci\'on se aprecia un efecto similar, al cerrarse una capa, el pr\'oximo nivel de energ\'ia disponible est\'a relativamente distante, y ello hace que se requiera una energ\'ia mucho menor para excitar el \'ultimo electr\'on (i.e. promoverlo a un nivel energ\'etico superior del espectro discreto) en un \'atomo alcalino que en un \'atomo de gas inerte. Ambos propiedades son las responsables del car\'acter qu\'imico distintivo de los elementos del Grupo IA y VIIIA.

Los gases nobles muestran una tendencia a formar compuestos comparativamente baja (de ah\'i gases ``inertes''). Sin embargo, los elementos que tienen un electr\'on m\'as o un electr\'on menos que los gases nobles son qu\'imicamente activos, y tienen propiedades contrastantes. Aquellos que siguen (en la Tabla Peri\'odica) a los gases nobles, alcalinos y alcalinos t\'erreos, como el litio, el berilio, el potasio, el calcio y otros, son met\'alicos y fuertemente electropositivos: tienden \textit{f\'acilmente} a formar iones positivos (baja energ\'ia de ionizaci\'on). Los elementos que se encuentran antes de los gases nobles, especialmente los hal\'ogenos, son el\'ectricamente no conductores, algunos incluso son gaseosos, y fuertemente electronegativos: tienden a formar iones negativos aislados o en combinaci\'on con otros \'atomos. Elementos de estos dos tipos suelen combinarse entre s\'i para formar compuestos estables; uno de los ejemplos m\'as comunes de estos compuestos es el cloruro de sodio ($NaCl$).

Si se asume que la actividad qu\'imica (i.e. la capacidad de formar mol\'eculas) est\'a de alguna manera condicionada a la magnitud de los campos externos el\'ectrico y magn\'etico creados por el \'atomo, entonces, en el caso de los gases nobles, estos campos deber\'ian ser considerablemente peque\~nos o nulos. 

Las propiedades de los elementos alcalinos pueden ser explicadas si se supone que el arreglo de los electrones externos en un \'atomo de gas noble es particularmente estable, o sea, una configuraci\'on de energ\'ia especialmente baja. En muchas ocasiones, para los \'atomos alcalinos, la p\'erdida de un electr\'on es favorable debido a que el sistema llega a una configuraci\'on electr\'onica similar a la de los gases nobles.

El estudio exhaustivo del sistema peri\'odico de los elementos y sus caracter\'isticas principales es abordado en los libros de texto de f\'isica at\'omica \cite{alonso} y \cite{richtmyer}, que se recomiendan al lector.

La presente tesis, como ya se ha dicho, aborda al estudio de la estabilidad de aniones at\'omicos en fase gaseosa (i.e. aislados), y, para ello, nos circunscribiremos al an\'alisis de dos grupos de \'atomos que tienen la estructura electr\'onica de los elementos del Grupo IA y del Grupo VIIIA de la Tabla Peri\'odica, o sea de los metales alcalinos y de los gases nobles, respectivamente. Ahora bien, debido a que la carga nuclear $Z$ se variar\'a entre $Z=N-2$ y $Z=N$ para construir las curvas de energ\'ia de ionizaci\'on, la identificaci\'on de estos grupos de \'atomos ($N=2,10,18,36,54,86$ y $N=3,11,19,37,55,87$) con los Grupos correspondientes de la Tabla Peri\'odica ($Z=2,10,18,36,54,86$ y $Z=3,11,19,37,55,87$, con $Z=N$) no es totalmente rigurosa. Expliquemos este detalle con un ejemplo; el caso de $N=18$ tiene un ani\'on con carga $Q=-1$ estable (\cite{11}) en fase gaseosa, de manera que tendr\'iamos un \'atomo con $N=18$ y $Z=17$ ($Cl^{-}$). En este caso, el arreglo de electrones es similar al del Arg\'on ($Ar$), pero un poco m\'as expandido debido a que la carga nuclear del $Ar$ es superior a la del $Cl$ ($Z_{Cl}+1=Z_{Ar}$). El $Cl^{-}$ tiene una estructura electr\'onica muy similar a la del $Ar$, aunque no id\'entica.

\begin{figure}[htbp]
  \includegraphics[width=\linewidth,angle=0]{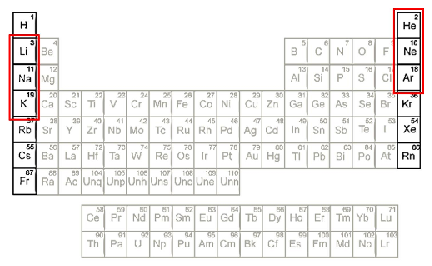}
  \caption{Tabla Peri\'odica de Mendel\'eev donde se han resaltado los Grupos IA y VIIIA, de estructura similar a los \'atomos estudiados. Se han enmarcado en cuadros rojos los elementos neutros con la misma cantidad de electrones que los \'atomos para los que fue calculada la energ\'ia de ionizaci\'on.}\label{periodic_table}
\end{figure}

Fueron escogidos tres \'atomos (v\'ease Fig. \ref{periodic_table}), de cada grupo se\~nalado, para realizar los c\'alculos num\'ericos de energ\'ia de ionizaci\'on: $N=2,10,18$, del primer grupo, y $N=3,11,19$, del segundo. Las comparaciones reportadas en la discusi\'on (Ep. \ref{stability}) muestran que la separaci\'on en grupos no es arbitraria.

%\newpage

\section{Modelo de una part\'icula para un \'atomo de $N$ electrones}
\label{modelo}

Para estudiar el estado d\'ebilmente ligado del \'ultimo electr\'on en un \'atomo de $N$ electrones se puede introducir un modelo semi-anal\'itico (\cite{1}, \cite{9}-\cite{11}) que describa las interacciones de ese electr\'on con el cori\'on (i.e. n\'ucleo at\'omico y electrones internos).

Es conocido que a larga distancia el electr\'on d\'ebilmente ligado s\'olo siente la atracci\'on de un n\'ucleo efectivo con carga $Z_{core}=Z-N+1$, o sea est\'a sujeto a un potencial del tipo
$-Z_{core}/r$. Sin embargo, a corta distancia pueden existir efectos de llenado de capas que pueden ser modelados por un pozo de potencial cuadrado finito con profundidad $-V_{0}$ y ancho $r_{0}$.

\begin{figure}[htbp]
  \includegraphics[width= .7 \linewidth, angle=0]{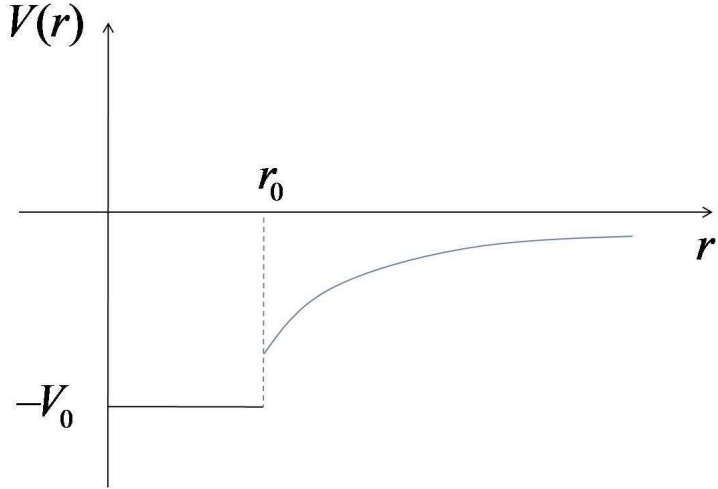}
  \caption{Esquema del modelo de potencial de una part\'icula $V(r)$ para $Z_{core}>0$. Se muestra un pozo cuadrado para $r \leq r_{0}$ y una cola de Coulomb para $r>r_{0}$.}\label{esquema}
\end{figure}

De manera que el potencial que se propone para el problema de una part\'icula puede escribirse de la siguiente manera:

\begin{equation}
V(r)=-\Theta(r,r_{0}) V_{0}-\Theta(r_{0},r) \frac{Z_{core}}{r}
\end{equation}

\noindent donde $\Theta(r,r')$ es la funci\'on de Heaviside que es nula para $r>r'$ e igual a $1$ para $r<r'$. En la Fig. \ref{esquema} aparece una representaci\'on esquem\'atica de este potencial.

Se debe resolver entonces, para este potencial, la ecuaci\'on:

\begin{equation}
\{ -\frac{1}{2}\nabla^{2} -\Theta(r,r_{0}) V_{0}-\Theta(r_{0},r)
\frac{Z_{core}}{r}\}\varphi(\vec{r})=\varepsilon \varphi(\vec{r})
\end{equation}

Planteando la separaci\'on de variables seg\'un la expresi\'on:

\begin{equation}
\varphi(\vec{r})=R(r)Y(\theta,\phi)
\end{equation}

\noindent donde $Y(\theta,\phi)$ son los arm\'onicos esf\'ericos, se obtiene la siguiente ecuaci\'on para la funci\'on de onda radial $R(r)$:

\begin{equation}
\{
\frac{1}{2}\frac{\partial^{2}}{\partial{r}^{2}}+\frac{1}{r}\frac{\partial}{\partial{r}}+
\Theta(r,r_{0}) V_{0}+\Theta(r_{0},r) \frac{Z_{core}}{r} -
\frac{l(l+1)}{r^{2}} + \varepsilon \}R(r)=0
\end{equation}

La funci\'on de onda que es soluci\'on de esta ecuaci\'on diferencial resulta:

\begin{eqnarray}\label{wavefunction_model}
R(r)=\bigg \{\begin{array}{cc}
          A \frac{J_{l+1/2}(\alpha' r)}{\sqrt{\alpha' r}} & r<r_{0} \\
         B \frac{W_{\beta,l+1/2}(\alpha r)}{r} & r>r_{0}
       \end{array}
\end{eqnarray}

\noindent donde:

\begin{equation}\label{betaalpha}
\left \{ \begin{array}{c}
  \beta = \frac{Z_{core}}{\sqrt{2 |\varepsilon|}} \\
  \alpha = \sqrt{8 |\varepsilon|} \\
  \alpha' = \sqrt{2 |\varepsilon|}
\end{array}
\right.
\end{equation}

\noindent $J_{l+1/2}(x)$ es la funci\'on de Bessel de orden fraccionario y $W_{\kappa,\mu}(x)$ es la soluci\'on de la Ecuaci\'on de Whittaker que es regular para $x\rightarrow \infty$. Para m\'as
informaci\'on sobre las propiedades de estas funciones cons\'ultese \cite{libro2}.

Para hallar los niveles de energ\'ia se utiliza la continuidad de la funci\'on de onda y su derivada en la frontera $r_{0}$. De manera que se obtiene la siguiente ecuaci\'on trascendente:

\begin{equation}
\alpha'(\varepsilon)\frac{J_{l+1/2}^{'}(\alpha'(\varepsilon)
r_{0})}{J_{l+1/2}(\alpha'(\varepsilon)
r_{0})}=\alpha(\varepsilon)\frac{W_{\beta,l+1/2}^{'}(\alpha(\varepsilon)
r_{0})}{W_{\beta, l+1/2}(\alpha(\varepsilon) r_{0})}
\end{equation}

\noindent cuya soluci\'on son los niveles de energ\'ia $\varepsilon$.

Este modelo se emplear\'a para deducir el comportamiento cerca del umbral de la energ\'ia de ionizaci\'on $E_{I}$ de un \'atomo con $N$ electrones, que ser\'ia la energ\'ia del estado b\'asico del electr\'on d\'ebilmente ligado $\varepsilon_{0}$.

%% file: cap_2.tex
\chapter{Formalismos te\'oricos y m\'etodos de la Qu\'imica Cu\'antica}
\markright{CAP\'ITULO \thechapter. FORMALISMOS TE\'ORICOS Y M\'ETODOS...}
\label{metodos}
\section{Hamiltoniano at\'omico}
\label{AtomHamiltoniano}

Consideremos un \'atomo o ion con carga nuclear $Z e$ y $N$ electrones. Un tratamiento detallado de este sistema debe tener en cuenta (\cite{libro1}): 1) la energ\'ia cin\'etica de los electrones y su energ\'ia potencial electrost\'atica (Coulomb) debida a la atracci\'on del n\'ucleo (asumiendo que es puntual y de masa infinita), 2) la repulsi\'on electr\'ostatica (Coulomb) entre electrones, 3)  la interacci\'on magn\'etica de los espines electr\'onicos con el orbital (interacci\'on esp\'in-\'orbita), y 4) otros efectos relativistas peque\~nos como la interacci\'on esp\'in-esp\'in, correcciones radiativas y nucleares (debido a la masa y dimensiones finitas del n\'ucleo, momento de dipolo magn\'etico nuclear, etc.).

Usualmente la descripci\'on detallada de un \'atomo de muchos electrones es sumamente dif\'icil y hay que recurrir a descripciones aproximadas. En este trabajo despreciamos las interacciones consideradas en 3) y 4). De manera que resulta el siguiente Hamiltoniano at\'omico (en unidades del Sistema Internacional):

\begin{equation}
\hat{H}=\sum_{i=1}^{N}{\{-\frac{\hbar^{2}}{2m_{e}}\nabla^{2}_{r_{i}}-(\frac{e^{2}}{4\pi\epsilon_{0}})\frac{Z}{r_{i}}\}}+\sum_{i<j=1}^{N}(\frac{e^{2}}{4\pi\epsilon_{0}})\frac{1}{r_{ij}}
\end{equation}

\noindent donde $r_{i}$ es la coordenada del electr\'on $i$ con respecto al n\'ucleo, $r_{ij} =|r_{i} - r_{j}|$ es la distancia entre el electr\'on $i$ y el $j$, y la \'ultima suma corre sobre
todos los pares de electrones.

N\'otese que este Hamiltoniano es independiente de las variables de esp\'in y es sim\'etrico ante permutaciones de las coordenadas espaciales de dos part\'iculas.

Es conveniente en este estudio pasar a unidades at\'omicas ($1$ hartree $=$ $27.211$ $eV$, $1$ bohr $=$ $0.52918$ ${\AA}$). El Hamiltoniano en estas unidades se escribe como:

\begin{equation}{\label{H_atom}}
\hat{H}=\sum_{i,j=1}^{N}{\{-\frac{1}{2}\nabla^{2}_{r_{i}}-\frac{Z}{r_{i}}\}}+\sum_{i<j=1}^{N}\frac{1}{r_{ij}}
\end{equation}

La Ecuaci\'on de Schr\"{o}dinger para la funci\'on de onda espacial
$\psi$ se escribe:

\begin{equation}
\hat{H}\psi(r_{1},\ldots,r_{N}) = E\psi(r_{1},\ldots,r_{N})
\end{equation}

\noindent o expl\'ictamente, usando la expresi\'on (\ref{H_atom}) de
$\hat{H}$:

\begin{equation}
[\sum_{i=1}^{N}{\{-\frac{1}{2}\nabla^{2}_{r_{i}}-\frac{Z}{r_{i}}\}}+\sum_{i<j=1}^{N}\frac{1}{r_{ij}}\psi(r_{1},\ldots,r_{N})]
= E\psi(r_{1},\ldots,r_{N})
\end{equation}

Esta es una ecuaci\'on diferencial en derivadas parciales de $3N$
dimensiones, donde las variables independientes son las coordenadas
espaciales de los electrones ( $r_{1}, r_{2}, ..., r_{N}$ ). Esta
ecuaci\'on no admite separaci\'on de variables debido al t\'ermino
$1/r_{ij}$, que da cuenta de la repulsi\'on mutua entre electrones.
Para el caso de dos electrones ese t\'ermino puede tratarse como
perturbaci\'on, pero en general ($N>2$) es muy grande y no es
aplicable el m\'etodo perturbativo.

En el Ep\'igrafe \ref{HF} se abordar\'an los m\'etodos de Hartree y Hartree-Fock,
basados ambos en el principio variacional lineal (Ep. \ref{variacional}), que
permiten solucionar de manera aproximada este problema.

\section{Principio variacional}
\label{variacional}

Sea $\langle \Phi|\hat{H}|\Phi \rangle$ el valor medio del
Hamiltoniano $\hat{H}$ sobre cierta funci\'on de onda $\Phi$. El
desarrollo de $\Phi$ en t\'erminos de las autofunciones de $\hat{H}$
es el siguiente:

\begin{equation}{\label{expansion}}
|\Phi\rangle = \sum_{n} C_{n} | \Psi_{n} \rangle
\end{equation}

Si se exige que $\Phi$ sea normalizada, como las funciones de $\hat{H}$ son ortonormales
($\langle\Psi_{n}|\Psi_{m}\rangle=\delta_{nm}$), se tiene que:

\begin{equation}
\langle\Phi|\Phi\rangle = \sum_{n} |C_{n}|^2 = 1
\end{equation}

Teniendo en cuenta estas expresiones y usando que
$\hat{H}\Psi_{n}=E_{n}\Psi_{n}$ se puede demostrar de manera
sencilla que la magnitud $E(\Phi)=\langle \Phi|\hat{H}|\Phi \rangle$
siempre es una cota superior para el menor autovalor de $\hat{H}$
(energ\'ia del estado b\'asico).

\begin{eqnarray}
\nonumber \langle \Phi|\hat{H}|\Phi \rangle & = & \sum_{m,n} C_{m}^{*}C_{n} \langle \Psi_{m}|\hat{H}|\Psi_{n} \rangle \\
\nonumber & = &\sum_{m,n} C_{m}^{*}C_{n} E_{n}\langle \Psi_{m}|\Psi_{n} \rangle \\
\nonumber & = & \sum_{m,n} C_{m}^{*}C_{n} E_{n} \delta_{mn} \\
\nonumber & = &\sum_{n} |C_{n}|^{2} E_{n} \geq E_{0} \sum_{n} |C_{n}|^{2} \\
& = &\sum_{n} |C_{n}|^{2} E_{n} \geq E_{0}
\end{eqnarray}

Esta simple relaci\'on es el Teorema Variacional, que m\'as
expl\'icitamente afirma que cualquier funci\'on de onda en el espacio de
Hilbert tiene una energ\'ia mayor que la energ\'ia del estado b\'asico.

Este teorema permite crear una estrategia para hallar una
aproximaci\'on a la energ\'ia y la funci\'on de onda del estado b\'asico
(\cite{21}). Por ejemplo, es posible proponer una familia de
funciones de onda normalizadas $\Phi_{\theta}$ dependiente del
par\'ametro $\theta$ y calcular el valor esperado de la energ\'ia
dependiente de $\theta$:

\begin{equation}{\label{theta}}
E(\theta)=\langle \Phi_{\theta}|\hat{H}|\Phi_{\theta} \rangle
\end{equation}

Entonces minimizando (\ref{theta}) se obtiene
la mejor aproximaci\'on a la energ\'ia del estado b\'asico, restringida a
la familia de funciones $\Phi_{\theta}$.

Otro posible m\'etodo consiste en desarrollar la funci\'on de onda en t\'erminos de cierta base de
funciones ortonormalizadas en el espacio de Hilbert, de manera similar a (\ref{expansion}):

\begin{equation}{\label{exp_1}}
|\Phi\rangle = \sum_{n} c_{n} | \psi_{n} \rangle
\end{equation}

En este caso las funciones $|\psi_{n}\rangle$ no son autofunciones de $\hat{H}$.

Luego se plantea el problema de extremos para el funcional:

\begin{equation}{\label{funcional}}
F(\{c_{n}\}) = \sum_{m,n} {c_{m}}^{*} c_{n} \langle
\psi_{m}|\hat{H}| \psi_{n} \rangle - \varepsilon (\sum_{n}
|c_{n}|^{2}-1)
\end{equation}

\noindent donde se ha introducido el multiplicador de Lagrange
$\varepsilon$ para tener en cuenta la restricci\'on (problema de
extremos condicionados) de que la funci\'on de onda debe ser
normalizada. La soluci\'on de este \'ultimo problema ser\'a dada en
t\'erminos de los coeficientes del desarrollo y el multiplicador de
Lagrange. Teniendo expl\'icitamente los coeficientes que minimizan
el funcional (\ref{funcional}), o sea que minimizan la expresi\'on:

\begin{equation}{\label{E(Phi)}}
E(\{c_{n}\})=\langle \Phi|\hat{H}|\Phi \rangle=\sum_{m,n} {c_{m}}^{*} c_{n} \langle \psi_{m}|\hat{H}| \psi_{n} \rangle
\end{equation}

\noindent dada la restricci\'on $\sum_{n} {|c_{n}|}^{2}=1$ es posible
calcular la energ\'ia del estado b\'asico simplemente sustituyendo en
(\ref{E(Phi)}).

El m\'etodo variacional es uno de los m\'as empleados en la Qu\'imica
Cu\'antica. En particular, el \'ultimo caso ser\'a \'util para deducir las
ecuaciones de Hartree-Fock e Interacci\'on de Configuraciones.

\section{M\'etodo de Hartree-Fock en \'atomos }
\label{HF}

En esta secci\'on propondremos dos m\'etodos para resolver el
problema de hallar la funci\'on de onda y la energ\'ia de \'atomos aislados planteado en el Ep\'igrafe \ref{AtomHamiltoniano}. Primeramente trataremos
el m\'etodo de Hartree, de manera breve, luego se describir\'ia el
m\'etodo de Hartree-Fock. Para una revisi\'on m\'as detallada del
tema se recomienda consultar \cite{20} y \cite{21}.

En la aproximaci\'on de Hartree, la m\'as simple, la funci\'on de onda se restringe a un producto de $N$ orbitales de un
electr\'on (producto de Hartree):

\begin{equation}
\Phi(r_{1},\ldots,r_{N})=\prod_{i}^{N}\phi_{i}(r_{i})
\end{equation}

\noindent donde no se toma en cuenta el esp\'in de los electrones ni la
antisimetr\'ia de la funci\'on de onda. Adem\'as se impone la condici\'on
adicional de que los orbitales de un electr\'on sean ortonormales.

Si se desarrollan los orbitales en t\'erminos de una base de funciones en el espacio de Hilbert, de la manera siguiente:

\begin{equation}\label{base}
\phi_{i}(r_{i})=\sum_{\alpha}c_{i\alpha} \varphi_{i\alpha}(r_{i})
\end{equation}

\noindent se puede escribir entonces para la funci\'on de onda del
sistema:

\begin{equation}
\Phi(r_{1},\ldots,r_{N})=\sum_{\alpha}\prod_{i}^{N} c_{i\alpha} \varphi_{i\alpha}(r_{i})
\end{equation}

Tomando:

\begin{eqnarray}
\nonumber c_{\alpha}&=&\prod_{i}^{N} c_{i\alpha} \\
f_{\alpha}(r_{i})&=&\prod_{i}^{N}\varphi_{i\alpha}(r_{i})
\end{eqnarray}

\noindent obtenemos la expresi\'on (\ref{exp_1}).

Construyendo el funcional (\ref{funcional}) y minimiz\'andolo se obtienen los coeficientes del desarrollo $\{c_{\alpha}\}$
que permiten hallar la funci\'on de onda y la energ\'ia del estado b\'asico (ve\'ase Ec. (\ref{E(Phi)}) ).

En el m\'etodo de Hartree-Fock (HF) s\'i se considera la
antisimetr\'ia de la funci\'on de onda del sistema de electrones (en
general, fermiones). Para ello se escribe la funci\'on de onda como
un determinante de Slater como sigue:

\begin{equation}\label{func_HF}
\Phi(x_{1},\ldots,x_{N})=\frac{1}{\sqrt{N!}}\left\vert
  \begin{array}{cccc}
    \phi_{1}(x_{1}) & \phi_{2}(x_{1}) & \ldots & \phi_{N}(x_{1}) \\
    \phi_{1}(x_{2}) & \phi_{2}(x_{2}) & \ldots & \phi_{N}(x_{2}) \\
    \vdots & \vdots & \ddots & \vdots \\
    \phi_{1}(x_{N}) & \phi_{2}(x_{N}) & \ldots & \phi_{N}(x_{N}) \\
  \end{array}
\right\vert
\end{equation}

\noindent donde la variable $x_{i}$ identifica a la coordenada
espacial del electr\'on i-\'esimo y a su proyecci\'on de esp\'in y
$\phi_{i}(x_{j})$ es el orbital electr\'onico i-\'esimo evaluado en
las coordenadas y proyecci\'on de esp\'in del electr\'on j-\'esimo.

Si se desarrolla a los orbitales en t\'erminos de una base de funciones en el espacio de Hilbert (\ref{base}), y se redefinen las funciones de la base del
sistema de part\'iculas y los coeficientes, an\'alogamente al caso anterior, se obtiene para la funci\'on de onda del
sistema una expresi\'on similar a (\ref{exp_1}).

Planteando el funcional $E(\Phi)=\langle \Phi |\hat{H}| \Phi \rangle$ se obtiene la ecuaci\'on:

\begin{eqnarray}\label{hf_ec}
E(\Phi)=\sum_{i}^{N}\langle \phi_{i} |\hat{T}| \phi_{i} \rangle + \frac{1}{2} \sum_{i,j: i \neq j}^{N} \{ \langle \phi_{i}\phi_{j}|\hat{V}|\phi_{i}\phi_{j} \rangle -\nonumber\\
 \langle \phi_{i}\phi_{j}|\hat{V}|\phi_{j}\phi_{i} \rangle \}
\end{eqnarray}

donde:

\begin{eqnarray}
\hat{T}(r_{i}) &=& -\frac{1}{2}\nabla^{2}_{r_{i}}-\frac{Z}{r_{i}} \\
\hat{V}(r_{i},r_{j}) &=& \frac{1}{|r_{i}-r_{j}|}
\end{eqnarray}

En la expresi\'on (\ref{hf_ec}) los t\'erminos $\langle
\phi_{i}\phi_{j}|\hat{V}|\phi_{i}\phi_{j} \rangle$ est\'an
relacionados con la interacci\'on directa (coulombiana) entre un par
de electrones $i,j$, mientras que los t\'erminos $\langle
\phi_{i}\phi_{j}|\hat{V}|\phi_{j}\phi_{i} \rangle$ se deben a la
interacci\'on de intercambio debida a la antisimetr\'ia de la
funci\'on de onda.

En el Ep. \ref{variacional} se demostr\'o que $E(\Phi)$ es una cota superior
para la energ\'ia del estado b\'asico de manera que minimizando el
funcional (\ref{hf_ec}) bajo la restricci\'on de que los orbitales
sean funciones normalizadas ($\langle \phi_{i}|\phi_{i}\rangle=1$)
se obtiene la mejor aproximaci\'on  a la energ\'ia y la funci\'on de
onda del estado b\'asico del sistema de part\'iculas. La ecuaci\'on
variacional a resolver ser\'a:

\begin{equation}\label{funcional_hf}
\delta E(\{\phi\}) - \sum_{i} \varepsilon_{i} \delta \langle
\phi_{i} | \phi_{i} \rangle=0
\end{equation}

Utilizando la expresi\'on (\ref{hf_ec}) se obtiene:

\begin{eqnarray}\label{mean_field}
\hat{T}| \phi_{i} \rangle + \{ \sum_{j}^{N} \langle
\phi_{j}|\hat{V}|\phi_{j} \rangle \} | \phi_{i} \rangle -
\sum_{i}^{N} \{ \langle \phi_{i}|\hat{V}|\phi_{j} \rangle |\phi_{i}
\} \\
\nonumber = \varepsilon_{i} | \phi_{i} \rangle
\end{eqnarray}

Esta ecuaci\'on puede transformarse en la siguiente:

\begin{eqnarray}\label{mean_field_0}
\{ \hat{T} +  \sum_{i}^{N} [ \hat{V}_{i}^{d} - \hat{V}_{i}^{ex} ] \}
|\phi_{i} \rangle = \varepsilon_{i} | \phi_{i} \rangle
\end{eqnarray}

\noindent donde $\hat{V}_{i}^{d}=\langle \phi_{i} |\hat{V}| \phi_{i}
\rangle$ y se define el operador $\hat{V}_{i}^{ex}$ seg\'un la
operaci\'on $\hat{V}_{i}^{ex} | f\rangle= \langle \phi_{i} |\hat{V}|
f \rangle | \phi_{i}\rangle$; o:

\begin{eqnarray}\label{mean_field_1}
\{ \hat{T} +  {\cal{V}}_{i}^{d} - {\cal{V}}_{i}^{ex} \} |\phi_{i}
\rangle = \varepsilon_{i} | \phi_{i} \rangle
\end{eqnarray}

\noindent donde ${\cal{V}}_{i}^{d}=\sum_{i}^{N}\hat{V}_{i}^{d}$ y
${\cal{V}}_{i}^{ex}=\sum_{i}^{N}\hat{V}_{i}^{ex}$.

Una caracter\'istica interesante de las ecuaciones de HF escritas en
su forma integro-diferencial (\ref{mean_field_1}) es que parecen ser
ecuaciones de Schr\"{o}dinger de una part\'icula con cierto
potencial $\hat{V} = {\cal{V}}_{i}^{d} - {\cal{V}}_{i}^{ex}$, o sea
que puede interpretarse como el movimiento de un electr\'on sujeto
al campo central coulombiano del n\'ucleo y a un potencial que da
cuenta de la interacci\'on aproximada con los restantes $N-1$
electrones (\cite{libro1}). Aunque, en rigor, las expresiones
(\ref{mean_field_1}) no son ecuaciones de autovalores como la Ec. de
Schr\"{o}dinger debido a que el potencial depende de los orbitales
$\{\phi\}$ a trav\'es de los operadores ${\cal{V}}_{i}^{d}$ y
${\cal{V}}_{i}^{ex}$.

El t\'ermino ${\cal{V}}_{i}^{d}$ representa el potencial promedio
debido a la presencia de los restantes $N-1$ electrones y el
t\'ermino ${\cal{V}}_{i}^{ex}$ da cuenta de los efectos de
intercambio entre el estado $|\phi_{i}\rangle$ y los restantes $N-1$
orbitales ocupados por electrones.

A continuaci\'on obtendremos una ecuaci\'on algebraica an\'aloga a la ecuaci\'on integro-diferencial \ref{mean_field_1} m\'as apropiada para la soluci\'on del problema mediante un algoritmo computacional. 

Sustituyendo en (\ref{hf_ec}) el desarrollo de los orbitales
$\{\phi\}$, similar a (\ref{base}), y haciendo un poco de algebra se
obtiene la ecuaci\'on:

\begin{eqnarray}
\nonumber && E(\Phi)=\sum_{i}^{N}\sum_{\alpha,\gamma}
C_{i\alpha}^{*}C_{i\gamma}\langle \varphi_{\alpha} |\hat{T}|
\varphi_{\beta} \rangle \\
\nonumber && + \frac{1}{2} \sum_{i,j: i \neq
j}^{N}\sum_{\alpha,\beta,\gamma,\delta}C_{i\alpha}^{*}C_{j\beta}^{*}C_{i\gamma}C_{j\delta}
\{ \langle
\varphi_{\alpha}\varphi_{\beta}|\hat{V}|\varphi_{\gamma}\varphi_{\delta}
\rangle - \langle
\varphi_{\alpha}\varphi_{\beta}|\hat{V}|\varphi_{\delta}\varphi_{\gamma}
\rangle \}\\
\end{eqnarray}

Minimizando el funcional $E(\Phi)=E(\{C_{i\alpha}\})$ respecto de los coeficientes del desarrollo de los orbitales
$C_{i\alpha}$, con la restricci\'on $\sum_{\alpha}|C_{i\alpha}|^2=1$ se obtienen las ecuaciones de HF en su forma algebraica:

\begin{equation}\label{hf_eq}
\sum_{\gamma} \{T_{\theta\gamma} +
\sum_{j}^{N}\sum_{\beta,\delta}C_{j\beta}^{*}C_{j\delta} \{ \langle
\theta \beta |\hat{V}|\gamma \delta \rangle - \langle \theta \beta
|\hat{V}|\delta \gamma \rangle \} \} C_{n\gamma}=\varepsilon_{n}
C_{n \theta}
\end{equation}

donde:

\begin{eqnarray}\label{coulomb}
\nonumber T_{\theta\gamma} &=& \langle \theta |\hat{T}| \gamma \rangle\\
\nonumber \langle \theta \beta |\hat{V}|\gamma \delta \rangle &=&\\
&=&\int\int \{\varphi_{\theta}^{*}(r_{1}) \varphi_{\beta}^{*}(r_{2})\frac{1}{|r_{1}-r_{2}|} \varphi_{\gamma}(r_{1}) \varphi_{\delta}(r_{2}) d^{3}r_{1} d^{3}r_{2}\} \delta_{S_{z_{\theta}} S_{z_{\gamma}}} \delta_{S_{z_{\beta}}S_{z_{\delta}}}
\end{eqnarray}

La soluci\'on de esta ecuaci\'on ser\'ian los coeficientes $C_{n\theta}$ y
las energ\'ias $\varepsilon_{n}$. Ahora bien, la ecuaci\'on
(\ref{hf_eq}) es no lineal y su soluci\'on no puede obtenerse de
manera algebraica. Sin embargo, puede resolverse por el m\'etodo
num\'erico autoconsistente. Reescribamos (\ref{hf_eq}) de la siguiente
manera:

\begin{equation}\label{autoconsistencia}
H[\{\bf{C}^{(n-1)}\}]\bf{C}^{(n)}=\varepsilon\bf{C}^{(n)}
\end{equation}

\noindent donde los elementos de matriz de $H[\{\bf{C}^{(n-1)}\}]$
se escriben como:

\begin{equation}
H_{\theta\gamma}[\{\textbf{C}^{(n-1)}\}] = T_{\theta\gamma} +
\sum_{j}\sum_{\beta,\delta}(C_{j\beta}^{(n-1)})^{*}C_{j\delta}^{(n-1)}
\{ \langle \theta \beta |\hat{V}|\gamma \delta \rangle - \langle
\theta \beta |\hat{V}|\delta \gamma \rangle \}
\end{equation}

N\'otese que este problema si se toma una aproximaci\'on inicial
$\textbf{C}^{(0)}$ arbitraria puede ser resuelto iterativamente
solucionando el problema de autovalores y autovectores
$H\textbf{C}=\varepsilon\textbf{C}$. En cada iteraci\'on se construye la
matriz $H$ a partir de los autovectores en la iteraci\'on anterior. 

Ahora bien, la dimensi\'on de la matriz $H$ y de los vectores $\textbf{C}$ (formados por la componentes $C_{ni}$ para $n$ fijo, o sea $\textbf{C}=(C_{n1},C_{n2},\ldots,C_{nN_{orb}})$) es, en general, mucho mayor que el n\'umero de electrones $N$; es igual al tama\~no de la base de una part\'icula $N_{orb}$ empleado (i.e. $1 < i < N_{orb}$). Por tanto, los autovectores $\textbf{C}$ correspondientes a los autovalores $\varepsilon$ que se utilizan en cada iteraci\'on, una vez ordenados los autovalores de menor a mayor, son los $N$ primeros; estos estados, que son los de menor energ\'ia, son los ocupados por electrones, de manera que la energ\'ia del estado b\'asico del sistema sea la m\'inima posible. 

Haciendo un poco de algebra se puede hallar una expresi\'on sencilla para la energ\'ia de HF (aproximaci\'on a la energ\'ia
del estado b\'asico seg\'un el m\'etodo de HF):

\begin{equation}\label{E_hf}
E_{HF}=\min_{\{\Psi\}}\{E(\Psi)\}=\frac{1}{2}\sum_{n}\{
\varepsilon_{n} + \sum_{\alpha,\beta} C_{n\alpha}^{*} C_{n\beta}
\langle \alpha |\hat{T}| \beta \rangle \}
\end{equation}

N\'otese que hemos partido del desarrollo de los orbitales de una
part\'icula en determinada base $\{\varphi\}$, o sea:

\begin{equation}
\phi_{i}(r)=\sum_{i}C_{i\alpha}\varphi_{\alpha}(r)
\end{equation}

La expresi\'on (\ref{E_hf}) ser\'a la energía exacta del sistema de
part\'iculas en presencia de un campo medio, o sea despreciando la
correlación debida a la interacción coulombiana entre electrones, si
la base de funciones de una part\'icula $\{\varphi\}$ es una base
completa (l\'imite de HF). Usualmente, para el cómputo, debe
truncarse la base de funciones y $E_{HF}$ es s\'olo una
aproximaci\'on. A medida que se aumente el tamaño de la base
$\{\varphi\}$, esta se aproximar\'a a ser completa y $E_{HF}$
converger\'a a la energ\'ia en el l\'imite de HF ($E_{HF_{limit}}$).

\section{Teorema de Koopmans. Energ\'ia de Ionizaci\'on y Electroafinidad.}

Bajo la aproximaci\'on de los orbitales ``congelados'', o sea asumiendo que los primeros $N-1$ orbitales son id\'enticos para los sistemas con $N$ y $N-1$ electrones, se puede obtener una aproximaci\'on para la energ\'ia de ionizaci\'on del sistema, i.e. la energ\'ia necesaria para eliminar un electr\'on del orbital $k$, usando la teor\'ia de HF (\cite{libro1}). Restando la ecuaci\'on (\ref{hf_ec}) para el sistema con $N-1$ part\'iculas de la misma ecuaci\'on para el sistema con $N$ part\'iculas, y utilizando la expresi\'on (\ref{mean_field}), se obtiene que $E_{I}=\Delta^{k}E_{HF}=E_{HF}(N)-E_{HF}^{k}(N-1)=\varepsilon_{k}$, donde $\varepsilon_{k}$ es la energ\'ia del k-\'esimo estado monoelectr\'onico ocupado (\cite{libro1},\cite{kohanoff},\cite{szabo}). Este resultado se conoce como teorema de Koopmans. De manera similar se puede calcular la electroafinidad, i.e. la energ\'ia necesaria para que una mol\'ecula neutra forme un ani\'on, y esta resulta $E_{A}=E_{HF}^{k}(N+1)-E_{HF}(N)=\varepsilon_{k}$, donde ahora el orbital k-\'esimo es un orbital desocupado (\cite{kohanoff}). 

Es pertinente precisar que la identificaci\'on de $\varepsilon_{k}$ con la energ\'ia de ionizaci\'on en el primer caso y con la electroafinidad en el segundo, no es rigurosa, debido a que se han ignorado los reajustes de los orbitales monoelectr\'onicos en los sistemas con $N-1$ y $N+1$ part\'iculas (\cite{libro1}). Los estados de $N \pm 1$ part\'iculas $\Psi^{N \pm 1}(x_{1},\ldots,x_{N \pm 1})$ (\ref{func_HF}), construidos a partir de los orbitales monoelectr\'onicos optimizados para el sistema de $N$ part\'iculas, no son los \'optimos para los sistemas con $N \pm 1$ part\'iculas (\cite{szabo}). Debe advertirse, adem\'as, que, a\'un siendo $E_{HF}(N)$ y $E_{HF}^{k}(N \pm 1)$ cotas superiores para la energ\'ia del estado b\'asico de los sistemas con $N$ y $N-1$ ($N+1$) electrones, la cantidad $\varepsilon_{k}$ no es una cota superior (inferior) para la energ\'ia de ionizaci\'on (electroafinidad) del sistema con $N$ electrones, debido a que se ha tomado la resta entre dos cotas superiores (\cite{libro1}). 

A pesar de que las energ\'ias de ionizaci\'on calculadas por este m\'etodo son bastante razonables dentro de la teor\'ia de HF, las electroafinidades no lo son, debido a que los estados desocupados o virtuales son incorrectos y normalmente no son estados ligados (\cite{kohanoff}).

\section{Correlaci\'on electr\'onica}
\label{corr}

Cuando se asume que el electr\'on se mueve en presencia del campo central del n\'ucleo y el campo promedio de los dem\'as electrones, es de esperar que haya errores provenientes de despreciar los efectos de la interacci\'on coulombiana instant\'anea entre electrones, debida al potencial $e^{2}/r_{12}$ (\cite{23}). A esta interacci\'on coulombiana instant\'anea se le suele denominar correlaci\'on electr\'onica. N\'otese que es independiente de los efectos de intercambio, debidos al Principio de Exclusi\'on de Pauli, por lo tanto ocurre entre cualquier par de electrones ya tengan espines paralelos o antiparalelos. El efecto fundamental de esta correlaci\'on es la disminuci\'on de la probalidad de encontrar dos electrones muy cerca y el correspondiente aumento de la probabilidad de encontrarlos suficientemente alejados uno del otro (\cite{corr_el}).

Como se ha se\~nalado en el ep\'igrafe anterior, el m\'etodo de HF, siendo una teor\'ia de campo medio, no tiene en cuenta la correlaci\'on electr\'onica, y como consecuencia, s\'olo puede hallar el $99\%$ de la energ\'ia total del estado b\'asico (\cite{corr_el}). Sin embargo, suele ser preciso en las predicciones de estructura molecular, no es muy costoso computacionalmente, y puede ser aplicado a sistemas grandes, e.g. mol\'eculas org\'anicas. Desafortunadamente, la parte de la energ\'ia que la aproximaci\'on de HF ignora tiene importantes efectos qu\'imicos, particularmente cuando los enlaces se forman o se rompen (\cite{corr_el}); en casos extremos como el de la mol\'ecula de $F_{2}$, el m\'etodo de HF es incapaz de describir el enlace(\cite{23}). 

La energ\'ia de correlaci\'on fue introducida por L\"{o}wdin (\cite{corr_el}) en la forma siguiente:

\begin{equation}
E_{corr}=E-E_{HF}
\end{equation}

\noindent donde $E$ es la energ\'ia exacta no relativista del sistema electr\'onico, y no puede ser hallada de manera experimental. Esta energ\'ia de correlaci\'on debe diferenciarse de la definida en el contexto de la Teor\'ia de Funcionales de Densidad (\textit{Density Functional Theory}; DFT) como la difrencia entre la energ\'ia exacta no relativista menos la suma de la energ\'ia cin\'etica, la energ\'ia directa de Coulomb, y la energ\'ia de intercambio; en este caso el operador de intercambio es local y por tanto la suma mencionada es diferente a la energ\'ia de HF (\cite{kohanoff}).

Dado que la funci\'on de onda de HF es el mejor determinante de Slater que puede obtenerse (aplicando el M\'etodo Variacional, v\'ease Ep. 1.2 y 1.3) cualquier enfoque que a\~nada la correlaci\'on tiene que relajar esta aproximaci\'on considerando m\'as de un determinante de Slater o una funci\'on de onda antisim\'etrica de otro tipo que describa al sistema de electrones (\cite{kohanoff}). En principio, cualquier funci\'on antisim\'etrica puede escribirse como una combinaci\'on lineal de infinitos determinantes de Slater (\cite{kohanoff}), aunque este no es un m\'etodo pr\'actico. Una mejor estrategia es incluir un peque\~no n\'umero de determinantes que sean realmente importantes en la energ\'ia (\cite{kohanoff}). La manera en que se construyen dichos determinantes se abordar\'a en el siguiente ep\'igrafe.

Es preciso apuntar que el cuadro de electrones individuales, representados por orbitales monoelectr\'onicos, en un campo medio de los dem\'as electrones (v\'ease Ep. 1.3) no se mantendr\'a en lo adelante. En general, la funci\'on de onda de muchas part\'iculas correlacionadas no puede ser escrita como un producto antisimetrizado de orbitales monoelectr\'onicos (\cite{kohanoff}). 

De cualquier modo, dado que de HF resulta la mayor contribuci\'on a la energ\'ia, la funci\'on de onda de HF es un buen punto de partida para c\'alculos m\'as elaborados.

\section{M\'etodo de Interacci\'on de Configuraciones en \'atomos}
\label{CI}

La Interacci\'on de Configuraciones (\textit{Configuration Interaction},
en lo adelante CI) es un m\'etodo que ha sido aplicado con \'exito a
estudios cuantitativos en sistemas electr\'onicos y en particular en
\'atomos (\cite{24}).

Es un m\'etodo post Hartree-Fock para encontrar la funci\'on de onda
y la energ\'ia del estado b\'asico y los estados excitados de un
sistema de muchas part\'iculas, teniendo en cuenta la correlaci\'on
entre ellas.

Se basa en el planteamiento del principio variacional lineal para la
funci\'on de onda del sistema, que se desarrolla en t\'erminos de una
base de funciones en el espacio de Fock de $N$ part\'iculas, de la
siguiente manera:

\begin{equation}\label{N_wavefunction}
\Psi^{(n)}=C_{0}^{(n)}\Phi_{0} +
\sum_{\sigma,\lambda}X_{\sigma\lambda}^{(n)} \Phi_{\lambda}^{\sigma}
+\sum_{\sigma,\tau,\lambda,\mu}Z_{\sigma\tau\lambda\mu}^{(n)}
\Phi_{\lambda\mu}^{\sigma\tau} + \hdots
\end{equation}

Usualmente las funciones de la base $\{\Phi\}$ suelen denominarse
configuraciones y tienen las mismas propiedades de simetr\'ia de la
funci\'on de onda del sistema de part\'iculas. En particular para el
caso de electrones (que son fermiones) esta es una base de
determinantes de Slater (funciones antisim\'etricas).

\begin{figure}[!h]
  \begin{center}
  \includegraphics[width=.9 \linewidth,angle=0]{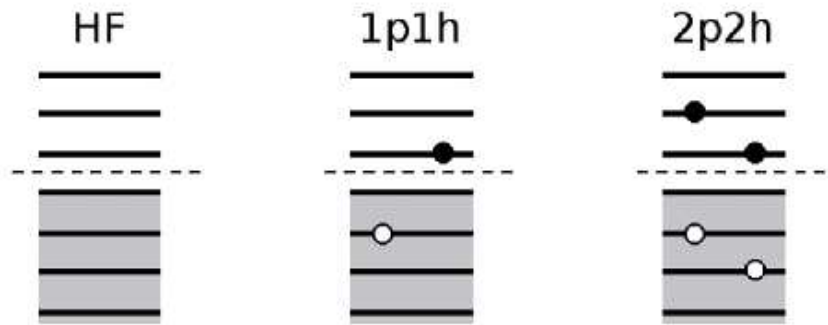}
  \end{center}
  \caption{Esquema que representa a las configuraciones HF, 1p1h, y 2p2h en el espectro electr\'onico de HF(l\'ineas continuas horizontales). Con fondo gris, se muestran los estados ocupados, y con fondo blanco, los desocupados. Los puntos negros son electrones y los blancos huecos.}\label{figura1}
\end{figure}

El determinante de Slater $\Phi_{0}$ , que aparece en el primer
t\'ermino, es la funci\'on de onda de HF. Los determinantes de
Slater subsecuentes
($\Phi_{\lambda}^{\sigma},\Phi_{\lambda\mu}^{\sigma\tau}, \hdots $)
se obtienen de sustituir orbitales ocupados $\lambda, \mu, \hdots$
(con energ\'ia menor que la energ\'ia de Fermi) por orbitales
virtuales o desocupados $\sigma, \tau, \hdots$(con energ\'ia mayor
que la energ\'ia de Fermi) de HF. O sea, los determinantes del
segundo t\'ermino son las configuraciones obtenidas a partir de
considerar una part\'icula en un orbital de HF desocupado $\sigma$ y
dejar vac\'io un orbital de HF ocupado $\lambda$, llamadas
configuraciones 1p1h (\textit{one-particle one-hole}) o excitaciones
simples; los del tercer t\'ermino son las configuraciones obtenidas
a partir de considerar dos part\'iculas en dos orbitales desocupados
$\sigma, \tau$  y dejar vac\'ios dos orbitales ocupados $\lambda,
\mu$, llamadas configuraciones 2p2h (\textit{two-particles
two-holes}) o excitaciones dobles (Fig. \ref{figura1}). Sustituyendo
$n$ orbitales ocupados por $n$ desocupados, se obtiene el t\'ermino
$n+1$ del desarrollo.

La magnitud $\langle\Phi|\hat{H}|\Phi\rangle$ siempre es mayor que
la energ\'ia exacta, de manera que minimiz\'andola respecto a los
coeficientes de los Slaters (principio variacional lineal, v\'ease
Ep. \ref{variacional}) se obtiene la mejor aproximaci\'on a la energ\'ia exacta.
Esto conduce a un problema de autovalores y autovectores (\cite{20}) que es posible representar (\cite{qdot_univ}), tomando hasta configuraciones 2p2h, como:

\begin{equation}\label{autoval}
\left(
  \begin{array}{ccc}
    E_{HF} & 0 & D^{t} \\
    0 & A & B^{t} \\
    D & B & C \\
  \end{array}
\right) \left(
  \begin{array}{c}
    C_{0} \\
    X \\
    Z \\
  \end{array}
\right)=E\left(
  \begin{array}{c}
    C_{0} \\
    X \\
    Z \\
  \end{array}
\right)
\end{equation}

\noindent donde $E_{HF}=\langle \Phi_{0} |\hat{H}|\Phi_{0} \rangle$
es la energ\'ia de HF, $D_{HF, \sigma\tau\lambda\mu}=\langle
\Phi_{0}|\hat{H}| \Phi_{\lambda\mu}^{\sigma\tau} \rangle$,
$A_{\sigma\lambda,\sigma' \lambda'}=\langle
\Phi_{\lambda}^{\sigma}|\hat{H}| \Phi_{\lambda'}^{\sigma'} \rangle$
es la matriz de Tamm-Dankoff,
$B_{\sigma\lambda,\sigma'\tau'\lambda'\mu'}=\langle
\Phi_{\lambda}^{\sigma}|\hat{H}|
\Phi_{\lambda'\mu'}^{\sigma'\lambda'} \rangle$, y
$C_{\sigma\tau\lambda\mu,\sigma'\tau'\lambda'\mu'}=\langle
\Phi_{\lambda\mu}^{\sigma\lambda}|\hat{H}|
\Phi_{\lambda'\mu'}^{\sigma'\lambda'} \rangle$.

\noindent (Las expresiones expl\'icitas de estos elementos de matriz se pueden consultar en el Ap\'endice \ref{appA}.).

En el problema (\ref{autoval}) los autovectores son el conjunto de
valores de los coeficientes $C_{0}, X_{\sigma\lambda},
Z_{\sigma\tau\lambda\mu}$ que hagan m\'inima la energ\'ia, y los
autovalores $E$, los niveles de energ\'ia para diferentes estados (el
estado b\'asico y los estados excitados).

Se puede demostrar de manera sencilla que tomando todas las
configuraciones posibles (FCI, \textit{Full Configuration
Interaction}), o sea una base completa de funciones $\{\Phi\}$, la
e\-ner\-g\'ia que resulta de este problema de autovalores es la energ\'ia
exacta (dentro del marco de la Mec\'anica Cu\'antica no-relativista), y
por ello es muy \'util como referencia para calibrar m\'etodos
aproximados (\cite{20},\cite{22}).

Ahora bien, como usualmente la base de funciones de una part\'icula
$\{\psi\}$ (orbitales) suele estar truncada (v\'ease \ref{HF}), la
energ\'ia obtenida con el m\'etodo FCI tampoco es la energ\'ia exacta.
Aunque es la mejor aproximaci\'on que se puede obtener para
determinado valor fijo de orbitales y el mismo tipo de base.

El desarrollo de la funci\'on de onda del sistema en la base de
funciones de $N$ part\'iculas ($\{\Phi\}$) tiene un gran n\'umero de
t\'erminos, incluso para una peque\~na base de orbitales
(\cite{22}). El Hamiltoniano (\ref{H_atom}), que no tiene en cuenta
las interacciones debidas al esp\'in, conmuta con los operadores de
proyecci\'on de esp\'in y esp\'in total ($\hat{S_{z}}$ y
$\hat{S^{2}}$), y por tanto los elementos de matriz $\hat{H}_{\mu
\nu}=\langle\Phi_{\mu}|\hat{H}|\Phi_{\nu}\rangle$ desaparecer\'an si
$\Phi_{\mu}$ y $\Phi_{\nu}$ son autofunciones de esp\'in con
diferentes autovalores $M_{s}$ o $S$. De manera que si la base de
funciones $\{\Phi\}$ son autofunciones de esp\'in se reduce
considerablemente la dimensi\'on de la matriz Hamiltoniana. A\'un
as\'i el n\'umero de t\'erminos de la base (Slaters, para
electrones) $\#_{config}$ aumenta dr\'asticamente con el n\'umero de
part\'iculas $N$ y el tama\~no de la base de orbitales $N_{orb}$:

\begin{equation}
\#_{config} = \sum_{n=0}^{N}
C_{n}^{N_{orb}-N}C_{n}^{N}=\sum_{n=0}^{N}\frac{(N_{orb}-N)!}{n!
(N_{orb}-N-n)!}\frac{N!}{(N-n)!n!}
\end{equation}

Para valores grandes de estas magnitudes el costo computacional de diagonalizar matrices de dimensi\'on grande hace que este m\'etodo (FCI) sea intratable. Por lo tanto es usual recurrir al truncamiento de la base de funciones de N part\'iculas hasta excitaciones simples (CIS), dobles (CISD), triples (CISDT), o cu\'adruples (CISDTQ). Debe tenerse en cuenta que, para buscar  una correcci\'on a la energ\'ia del estado b\'asico, la base de funciones debe extenderse al menos hasta excitaciones dobles. Esto \'ultimo es consecuencia de que el estado de HF no se mezcla con los estados 1p1h ($\langle \Phi_{0}|\hat{H}|\Phi_{\lambda}^{\sigma}\rangle=0$), Teorema de Brillouin (\cite{szabo}). Estos m\'etodos (CISD, CISDT, CISDTQ) tienen en cuenta una parte de la correlaci\'on electr\'onica, no toda, y se mejorar\'a la aproximaci\'on en tanto crezca el tama\~no de la base de $N$ part\'iculas considerada, o sea en tanto las funciones $\{\Phi\}$ se aproximen a una base completa.

El m\'etodo FCI es consistente al tama\~no (\textit{size-consistent}), o
sea la energ\'ia de dos sistemas suficientemente lejanos es la suma de
las energ\'ias de cada uno de los sistemas independientemente,  y es
extensivo (\textit{size-extensive}), o sea la energ\'ia del sistema es
una funci\'on lineal del n\'umero de part\'iculas $N$. Esto no ocurre con
los m\'etodos CI truncados (\cite{22}, \cite{23}).

%% file: cap_3.tex
\chapter{Resultados y discusi\'on}
\label{resultados}

\section{Implementaci\'on y detalles computacionales}
\label{comp}

El m\'etodo de HF fue implementado para \'atomos en FORTRAN 90, en forma de m\'odulo. Algunas peculiaridades de la implementaci\'on se detallan m\'as adelante. Todos los c\'alculos de CISD fueron ejecutados utilizando una adaptaci\'on para \'atomos de la librer\'ia QDOTLIB (\cite{qdotlib}), originalmente concebida para el c\'alculo de propiedades electr\'onicas de puntos cu\'anticos. El m\'odulo de HF y la librer\'ia QDOTLIB emplean subrutinas de la librer\'ia de \'Algebra Lineal LAPACK de la librer\'ia num\'erica MKL. Para diagonalizar matrices muy grandes (de dimensi\'on mayor o igual que $20000$) se utiliza el algoritmo de Lanczos (v\'ease Ap\'endice \ref{appB}).

La base de funciones de una part\'icula empleada es la base de \'atomos hidrogenoides:

\begin{equation}\label{hidrogenoid}
\Psi_{nlm}(\vec{r})=R_{nl}(r)Y_{lm}(\theta,\phi)
\end{equation}

\noindent donde $Y_{lm}(\theta,\phi)$ son los arm\'onicos esf\'ericos y la expresi\'on de $R_{nl}(r)$ normalizada viene dada por:

\begin{equation}
R_{nl}(r)=-\left \{ \left ( \frac{2Z}{na_{0}} \right )^{3} \frac{(n-l-1)!}{2n[(n+l)!]^{3}} \right \}^{1/2} e^{-\rho/2} \rho^{l} L_{n+l}^{2l+1}(\rho)\\
\end{equation}

\noindent con:

\begin{eqnarray}
\rho &=& \frac{2Z}{na_{0}} r \\
a_{0} &=& \frac{4 \pi \epsilon_{0} \hbar^{2}}{m {e}^{2}}
\end{eqnarray} 

\begin{figure}[htb]
  \begin{center}
  \includegraphics[width=.6\linewidth,angle=0]{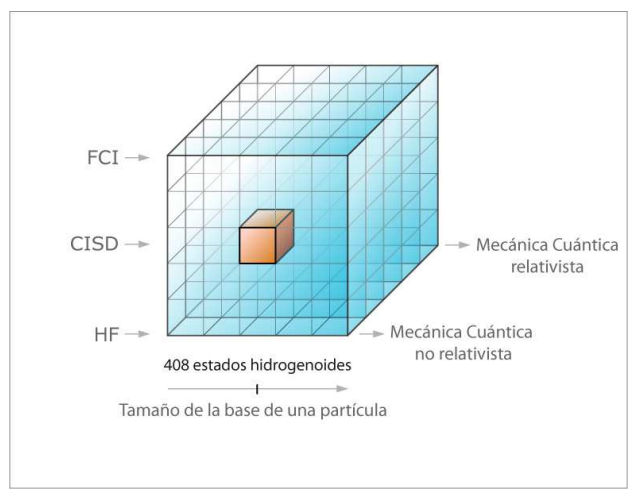}
  \end{center}
  \caption{Esquema en que se registran las aproximaciones en el c\'alculo \textit{ab initio} implementado. En el eje horizontal, en profundidad, se se\~nala si el Hamiltoniano tiene en cuenta alg\'un efecto relativista; en este estudio se trabaja dentro de la Mec\'anica Cu\'antica no relativista. En el eje horizontal, de izquierda a derecha, se indica el tama\~no de la base de una part\'icula (orbitales); 408 estados hidrogenoides en este caso. Y, en el eje vertical, se muestra el tipo de base de $N$ part\'iculas considerada; hasta excitaciones dobles en nuestro caso.}\label{cubic}
\end{figure}

\noindent En particular fueron utilizados s\'olo $408$ estados de esta base (Fig. \ref{cubic}), desde el $1s$($n=1,l=0$) hasta el $8j$ ($n=8, l=7$), $204$ con proyecci\'on de esp\'in $s_{z}=1/2$ y $204$ con $s_{z}=-1/2$.

Con esta base fueron calculados, previamente, los elementos de
matriz del potencial de Coulomb $\langle \varphi_{\theta} \varphi_{\beta} |1/|\vec{r}_{i}-\vec{r}_{j}| \varphi_{\gamma} \varphi_{\delta} \rangle$ (v\'ease Ap\'endice \ref{appC}). Los \'indices $\theta$, $\beta$, $\gamma$ y $\delta$ son dependientes de los n\'umeros cu\'anticos $n$, $l$, $m$, $s_{z}$. Las reglas de selecci\'on para estas integrales bielectr\'onicas permiten que no se calculen y no se almacenen los elementos matriciales de Coulomb
nulos.

El m\'etodo de HF implementado tiene las siguientes particularidades:

1. Se desprecian los coeficientes que son menores que cierto valor arbitrario (en particular, en los c\'alculos presentados se escogi\'o el valor $0.003$), i.e. se eval\'uan como nulos si cumplen dicha desigualdad.

2. El criterio de parada (condici\'on de convergencia) que se impuso fue el siguiente:

\begin{equation}
|\varepsilon_{n}^{(i)}-\varepsilon_{n}^{(i-1)}|<10^{-6}; \forall
\varepsilon_{n}<\varepsilon_{f}
\end{equation}

donde $\varepsilon_{f}$ es la energ\'ia de Fermi.

3. Adem\'as, en la autoconsistencia se usa el m\'etodo de subrelajaci\'on para garantizar la convergencia, de manera que la matriz Hamiltoniana diagonalizada en la iteraci\'on i-\'esima no es
precisamente H sino:

\begin{equation}
H^{(i)}=(1-\omega)H^{(i-1)}+\omega H(\{\bf{C}^{(i)}\})
\end{equation}

\noindent donde $\omega$ es un par\'ametro de fricci\'on o historia que est\'a entre $0$ y $1$. En la gran mayor\'ia de los c\'alculos que se presentar\'an en este art\'iculo se utiliz\'o $\omega = 0.3$ aunque en el caso de los c\'alculos para $N=2$ se requiri\'o un valor $\omega = 0.5$.

Adem\'as, se implement\'o el m\'etodo Post-Hartree-Fock de CISD (Fig. \ref{cubic}) con un truncamiento de la base de funciones de $N$ part\'iculas por energ\'ia y por reglas de selecci\'on por proyecci\'on de momento
angular y de esp\'in, adem\'as de por tipo de excitaciones consideradas (1p1h y 2p2h). O sea, no se consideran los Slaters $\Phi_{\lambda}^{\sigma}$ cuando:

\begin{eqnarray}\label{1p1h_selec_rules}
\nonumber |\varepsilon_{\lambda}-\varepsilon_{\sigma}|>E_{corte} \\
|m_{\lambda}-m_{\sigma}| \neq \triangle{m} \\
\nonumber |{s_{z}}_{\lambda}-{s_{z}}_{\sigma}| \neq \triangle{s_{z}}
\end{eqnarray}

y los Slaters $\Phi_{\lambda\mu}^{\sigma\tau}$ cuando:

\begin{eqnarray}\label{2p2h_selec_rules}
\nonumber |\varepsilon_{\lambda}+\varepsilon_{\mu}-\varepsilon_{\sigma}-\varepsilon_{\tau}|>E_{corte} \\
|m_{\lambda}+m_{\mu}-m_{\sigma}-m_{\tau}| \neq \triangle{m} \\
\nonumber
|{s_{z}}_{\lambda}+{s_{z}}_{\mu}-{s_{z}}_{\sigma}-{s_{z}}_{\tau}|
\neq \triangle{s_{z}}
\end{eqnarray}

Se escogieron los valores $\triangle{m}=0$ y $\triangle{s_{z}}=0$ para las dos \'ultimas condiciones. Se utiliz\'o el valor de $E_{corte}$ tal que $\#_{config}$ fuese aproximadamente igual a $5000$ para el caso $N=2$, $11000$ para $N=3$, y $50000$  para $N=10,11,18,19$.

Se hicieron c\'alculos de HF y CISD para para sistemas at\'omicos con $N = 2, 3, 10, 11, 18, 19$ y para sistemas con un electr\'on menos en cada caso (salvo en el caso en que $N-1=1$, cuyo resultado
es anal\'itico y muy bien conocido) para $Z_{core} \epsilon (-1,1)$ con paso $\Delta Z_{core}=0.1$. El objetivo de estos c\'alculos es hallar para cada $Z_{core}$ la energ\'ia de ionizaci\'on
$E_{I}(Z_{core})=E(N,Z_{core})-E(N-1,Z_{core})$.

%%%%%%%%%%%%%%%%%%%

Los c\'alculos de la energ\'ia del estado b\'asico de los sistemas at\'omicos con $N$ y $N-1$ electrones son aproximados debido al truncamiento de la bases de una part\'icula y de $N$ part\'iculas. Sin embargo, dado un tama\~no fijo de la base de una part\'icula (en nuestro caso $408$ estados de la base de \'atomos hidrogenoides), la mejor aproximaci\'on a la energ\'ia del estado b\'asico se obtiene incluyendo m\'as efectos de correlaci\'on electr\'onica, o sea aumentando el tama\~no de la base de $N$ part\'iculas $\#_{config}$; esta aproximaci\'on debe converger cuando $\#_{config}$ sea suficientemente grande,  i.e. cuando se toman en cuenta todas las configuraciones posibles (\textit{Full Configuration Interaction}) en el desarrollo de la funci\'on de onda del sistema (\ref{N_wavefunction}).

\begin{figure}[htbp]
  \begin{center}
  \includegraphics[width=.9\linewidth,angle=0]{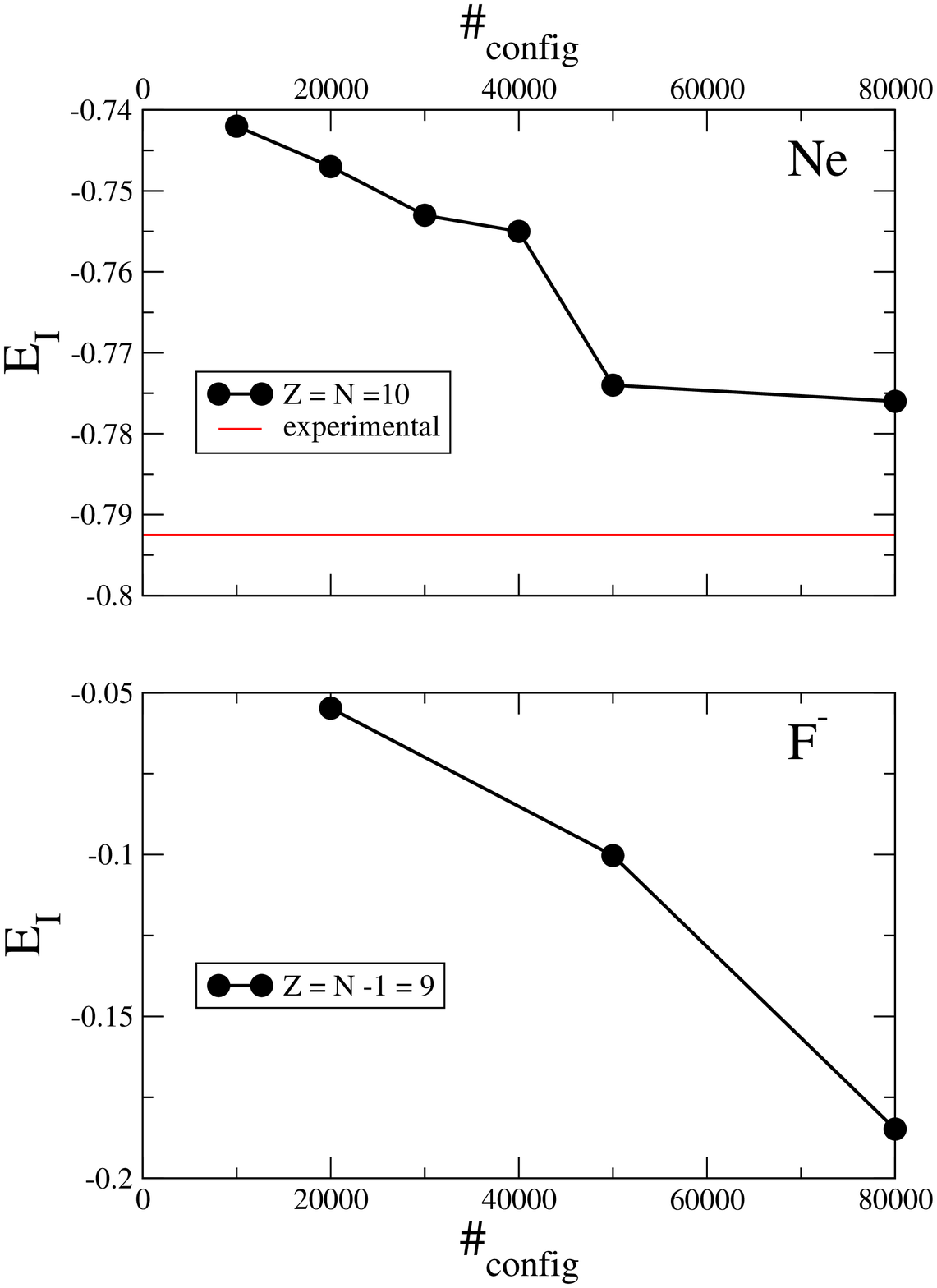}
  \end{center}
  \caption{Gr\'afico de $E_{I}$ vs. $\#_{config}$ para el caso $Ne$ y $F^{-}$. La l\'inea roja representa la energ\'ia de ionizaci\'on del $Ne$ obtenida experimentalmente.}\label{convergence}
\end{figure}

Un estudio de la dependencia de $E_{I}$ con $\#_{config}$ se muestra en la Fig. \ref{convergence} para el caso de $Z=10$ y $N=10$, \'atomo neutro de Ne\'on ($Ne$), y para el caso de $N=10$ y $Z=9$, ani\'on simplemente cargado de Fl\'uor ($F^{-}$). En el gr\'afico $\#_{config}$ es tal que no incluyen configuraciones m\'as all\'a de las excitaciones dobles (CISD).

N\'otese, ante todo, que la convergencia de $E_{I}$ no se obtiene en ninguno de los dos casos, a\'un cuando $\#_{config}=80000$. En el caso del $Ne$, se pueden comparar los valores de $E_{I}$ calculados con el obtenido de forma experimental \footnote{Todos los resultados experimentales considerados provienen de la base de datos del NIST (\textit{National Institute of Standards and Technology}), cuya direcci\'on electr\'onica es \textbf{www.nist.gov} .} (l\'inea roja). Adem\'as, podemos advertir comparando el comportamiento para $F^{-}$ y para $Ne$ que la convergencia no se obtendr\'a para el mismo n\'umero de configuraciones. 

De manera que la energ\'ia de ionizaci\'on calculada num\'ericamente no es una aproximaci\'on \textit{buena} a la energ\'ia de ionizaci\'on del sistema f\'isico real. No obstante, el pr\'oposito de este trabajo no es obtener con exactitud la energ\'ia de ionizaci\'on de los \'atomos estudiados sino describir caracter\'isticas generales del comportamiento de esta magnitud en el dominio ani\'onico para dichos sistemas, y, para ello, los c\'alculos num\'ericos ejecutados resultan claramente satisfactorios, como veremos m\'as adelante.

Existe cierta indefinici\'on acerca de c\'omo calcular la energ\'ia de ionizaci\'on de un \'atomo con $N$ electrones utilizando el algoritmo de Interacci\'on de Configuraciones truncado, y uno puede plantearse dos variantes: (1) calcular la energ\'ia del estado b\'asico de los \'atomos con $N$ y $N-1$ electrones para un n\'umero de configuraciones fijo ($\#_{config}$), aproximadamente igual en los dos sistemas, escogiendo diferentes energ\'ias de corte ($E_{corte}$) tales que verifiquen en cada caso que se tiene el valor correcto de $\#_{config}$; o (2) calcular la energ\'ia del estado b\'asico de ambos sistemas para la misma $E_{corte}$, o sea con un n\'umero de configuraciones diferente en cada caso. El an\'alisis de la dependencia de $E_{I}$ con $\#_{config}$ y las comparaciones con resultados experimentales, en cada caso, permitieron concluir que el esquema de c\'alculo m\'as adecuado es el (2). (En el caso en que se fija $E_{corte}$, el n\'umero de configuraciones al que se hace referencia es el del sistema con $N$ part\'iculas.). De manera que todos los c\'alculos, incluso los mostrados en la Fig. \ref{convergence}, fueron realizados fijando, para ambos sistemas, el valor de $E_{corte}$, que como ya se ha visto es el par\'ametro que limita las excitaciones consideradas en la construcci\'on de los Slaters. 

\section{Espectros de energ\'ia de \'atomos en el dominio \\ ani\'onico}
\label{spectra}

Una de las formas m\'as elementales de entender la estabilidad del \'atomo de $N$ electrones para $Z<N$ es analizar qu\'e le ocurre a su espectro energ\'etico a medida que se disminuye la carga nuclear. 

Para un an\'alisis fenomenol\'ogico de esta situaci\'on utilicemos el modelo de una part\'icula enunciado en \ref{modelo}. En el caso neutro, $Z_{core}>0$, el espectro energ\'etico del electr\'on m\'as lejano al n\'ucleo tiene infinitos niveles, y estos tienen un punto de acumulaci\'on (\textit{continuum}) para $\varepsilon=0$, debido a que el electr\'on considerado, a largas distancias, s\'olo percibe un n\'ucleo con carga efectiva $Z_{core}$ (\'atomo hidrogenoide con carga nuclear $Z_{core}$). Ahora bien, cuando $Z_{core}=0$ se tiene que la cantidad de niveles de energ\'ia para en el modelo de un part\'icula es finita, debido a que la profundidad del pozo cuadrado, utilizado para describir la interacciones de corto alcance, es finita. De manera que el espectro energ\'etico colapsa, i.e. pasa de una cantidad infinita a una cantidad finita de niveles de energ\'ia, en $Z_{core}=0$. Para $Z_{core}<0$ la cantidad de niveles debe decrecer con la disminuci\'on de la carga nuclear y, finalmente, cuando $Z_{core}=Z_{core}^{(c)}$ el estado b\'asico del electr\'on m\'as lejano entra en el \textit{continuum} y el sistema con $N$ part\'iculas se autoioniza.

Los c\'alculos num\'ericos de la energ\'ia del estado b\'asico y los primeros estados excitados de un sistema con $N$ electrones tambi\'en permiten visualizar este comportamiento. Si definimos la energ\'ia de ionizaci\'on $E_{I}^{n_{s}}$ como la energ\'ia necesaria para mantener ligado al electr\'on m\'as lejano al n\'ucleo, encontr\'andose el sistema en el estado $n_{s}$, se tiene que $E_{I}^{n_{s}}=E^{n_{s}}(Z,N)-E^{1}(Z,N-1)$, donde $E^{n_{s}}(Z,N)$ es la energ\'ia del estado $n_{s}$ del sistema con $N$ electrones y carga nuclear $Z$. En la Fig. \ref{spectra} se muestra el espectro $E_{I}^{n_{s}}$ para el dominio ani\'onico ($Z_{core} \leq 1$) en el caso de $N=10$, donde se muestran solamente aquellos estados $n_{s}$ para los cuales $E_{I}^{n_{s}}<0$ en la regi\'on del \'atomo neutro ($Z_{core}=1$).

\begin{figure}[htbp]
  \begin{center}
  \includegraphics[width=.7\linewidth,angle=0]{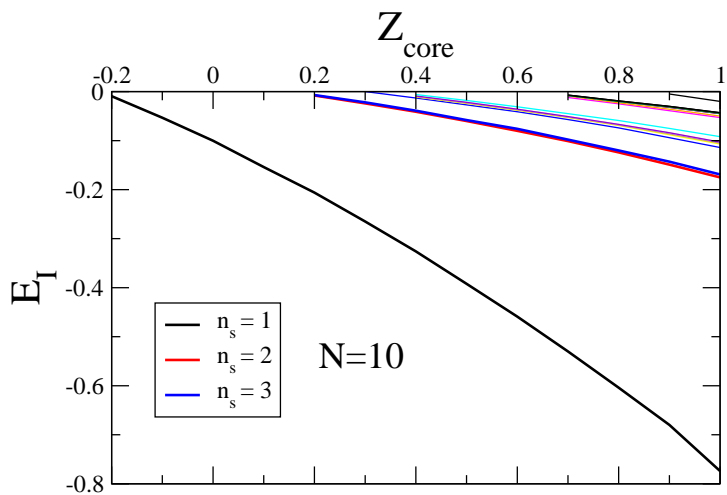}
  \end{center}
  \caption{Espectro energ\'etico de energ\'ias de ionizaci\'on del sistema con $N=10$ para $Z_{core} \leq 1$. En la figura se muestran los primeros $22$ estados para los cuales $E_{I}<0$.}\label{spectra}
\end{figure}

Obs\'ervese que por el contrario de la descripci\'on fenomenol\'ogica dada, la cantidad de niveles energ\'eticos es finita en todo el intervalo de $Z_{core}$. Esto se debe al truncamiento de la base de una part\'icula. Adem\'as, la raz\'on por la cual no aparecen m\'as estados para determinado $Z_{core}$ puede no estar relacionada con propiedades intr\'insecas del sistema, y s\'i con los orbitales monoelectr\'onicos escogidos; en espec\'ifico, es posible que se deba a que las funciones de onda consideradas en la base tengan un radio caracter\'istico menor que las necesarias para describir la interacci\'on de largo alcance del \'ultimo electr\'on con el n\'ucleo efectivo cuando la carga nuclear es suficientemente peque\~na. Sin embargo, lo m\'as importante que se debe resaltar de la Fig. \ref{spectra}, es que, en efecto, el estado b\'asico y los estados excitados van aproxim\'andose a $E_{I}=0$ cuando disminuye $Z_{core}$, hasta que, para $Z_{core}=Z_{core}^{(c)}$, todos los estados son absorbidos por el \textit{continuum} y el sistema se vuelve inestable. (V\'ease `absorci\'on de autovalores' en sistemas d\'ebilmente ligados en \cite{13}-\cite{15}.).

\section{Estabilidad y universalidad de aniones at\'omicos en fase gaseosa}
\label{stability}

Los c\'alculos num\'ericos de CI para la energ\'ia de ionizaci\'on de \'atomos con $N$ electrones arrojaron que para la carga nuclear cr\'itica se cumple que:

\begin{equation}
N-2 < Z^{(c)} < N-1
\end{equation}

\noindent para los \'atomos con la estructura electr\'onica de gases
nobles ($N=2,10,18$), v\'ease la Tabla \ref{tabla_zc}. Lo que indica
que, en fase gaseosa, existen aniones at\'omicos simples ($Z=N-1$)
estables, y que los aniones at\'omicos doblemente cargados ($Z=N-2$)
son inestables, resultado que est\'a en correspondencia con lo
obtenido por Kais y Serra (\cite{11}).

\begin{table}[h]
\begin{center}
\begin{tabular}{|c|ccc|}
\hline
$N$ & $2$ & $10$ & $18$ \\ \hline
$Z_{core}^{(c)}$ & $-0.035246$ & $-0.219341$ & $-0.353622$\\
\hline
\end{tabular}
\caption{\label{tabla_zc} Tabla de $Z_{core}^{(c)}$ para $N=2, 10,
18$.}
\end{center}
\end{table}

Sin embargo, en el caso de los metales alcalinos ($N=3,11,19$) se obtuvo $Z^{(c)} \geq N-1$,
de manera que no existen los aniones at\'omicos simples con la
estructura electr\'onica de los metales alcalinos.

\begin{figure}[htbp]
  \begin{center}
  \includegraphics[width=.9\linewidth,angle=0]{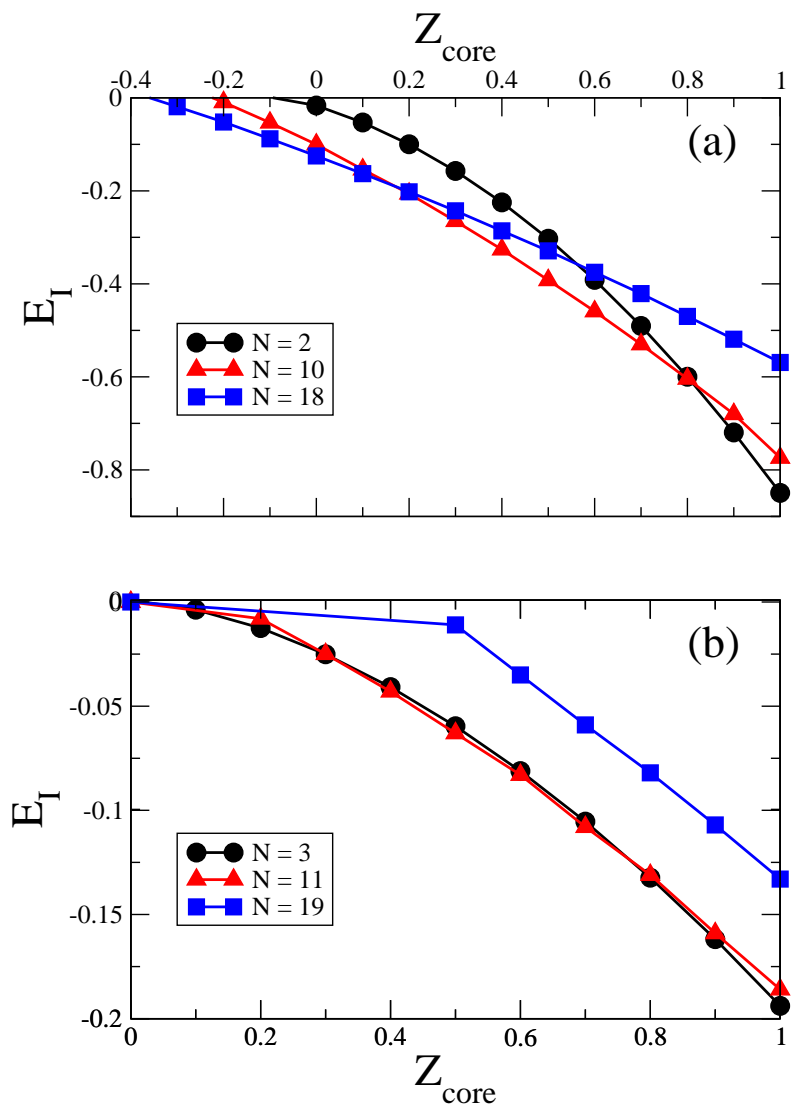}
  \end{center}
  \caption{Gr\'afico de $E_{I}$ vs. $Z_{core}$. En (a) se muestran los casos de \'atomos con la estructura electr\'onica de gases nobles y en (b) los casos con la estructura electr\'onica de los metales alcalinos. Obs\'ervese que $Z_{core}^{(c)}<0$, es decir $Z^{(c)}<N-1$, para los gases nobles.}\label{figura3}
\end{figure}

Otra caracter\'istica distintiva de estos dos grupos de \'atomos es
el comportamiento de $E_{I}$ en la vecindad
del valor umbral de $Z_{core}$. En la Fig. \ref{figura3} se muestra
dicha diferencia.

N\'otese primero que en el caso de los gases nobles ocurre que los
elementos con mayor n\'umero de electrones son menos estables en la
regi\'on del \'atomo neutro, sin embargo a partir de cierto valor de
$Z_{core}$ esta caracter\'istica se invierte, i.e. las curvas se
cruzan, y en la regi\'on ani\'onica cercana al valor umbral
$Z_{core}=Z_{core}^{(c)}$ son m\'as estables los \'atomos con m\'as
electrones (v\'ease tambi\'en la Tabla \ref{tabla_zc}).

En el caso de los metales alcalinos se muestra que este
caracter\'istica no est\'a presente, i.e. los elementos con menor
n\'umero de electrones son siempre m\'as estables. Es preciso hacer
notar que para $N=11$ y $N=19$ los puntos con $Z_{core}<0.2$ y
$Z_{core}<0.5$, respectivamente, no fueron calculados correctamente
con los m\'etodos num\'ericos utilizados, y que la uni\'on con el
punto $(0,0)$ es simplemente una gu\'ia visual. Este c\'alculo puede
mejorarse aumentando la cantidad de configuraciones consideradas en
la funci\'on de onda de CI (\ref{N_wavefunction}) y aumentando la
cantidad de orbitales de una part\'icula utilizados en HF.

Para hallar una expresi\'on anal\'itica para el comportamiento de
$E_{I}$ en las proximidades del valor umbral $Z_{core}^{(c)}$, en
ambos casos, utilizamos el modelo propuesto en el ep\'igrafe \ref{modelo}.
Previamente se hizo un ajuste de los c\'alculos num\'ericos con los
resultados del modelo para el primer nivel de energ\'ia, energ\'ia
de enlace. Este ajuste no es perfecto pero para valores cercanos a
$Z_{core}^{(c)}$ es muy bueno, de manera que este modelo es \'util
para estudiar el comportamiento anal\'itico en la regi\'on umbral.

\begin{figure}[h]
  \begin{center}
  \includegraphics[width= .7 \linewidth,angle=0]{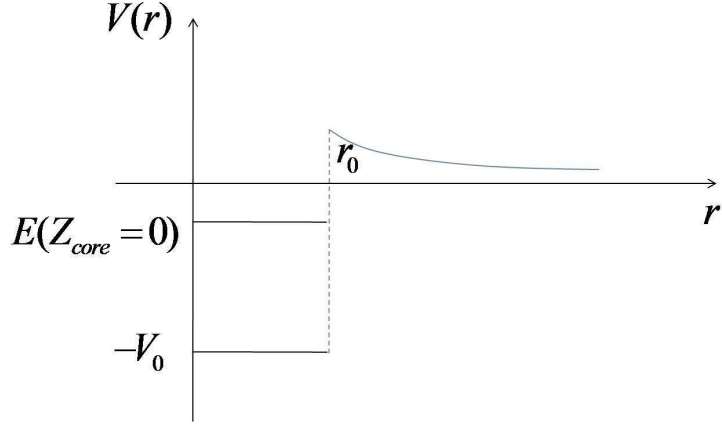}
  \end{center}
  \caption{Esquema que representa el potencial de una part\'icula caracter\'istico para los \'atomos con capas electr\'onicas
  llenas.}\label{esquema_gases_nobles}
\end{figure}

El caso de los gases nobles puede ser modelado con el potencial de
una part\'icula del Ep. \ref{modelo}, de manera que $V(r)$ sea tal que
admita al menos un estado ligado para $Z_{core}=0$, o sea que el
pozo cuadrado tenga la suficiente profundidad $V_{0}$ como para
admitir un nivel de energ\'ia. Ahora bien, la zona de inter\'es en
el dominio de $Z_{core}$ ser\'a $Z_{core}<0$, donde el potencial de
Coulomb es repulsivo. Para esta regi\'on debe existir un valor
cr\'itico $Z_{core}=Z_{core}^{(c)}$ para el cual el electr\'on m\'as
lejano al n\'ucleo deje de estar ligado, i.e. se produzca efecto
t\'unel a trav\'es de la barrera coulombiana repulsiva (v\'ease Fig.
\ref{esquema_gases_nobles}).

Para problemas de dos part\'iculas (e.g. el cori\'on y un electr\'on) con
un potencial repulsivo menor que decae m\'as lentamente que
$\frac{1}{r^{2}}$ (en particular el potencial coulombiano
$|Z_{core}|/r$) es conocido (\cite{13}) que la funci\'on de onda en
el estado de cero energ\'ia del umbral tiene radio finito, de manera
que en la aproximaci\'on a la zona umbral no hay un cambio brusco en
la funci\'on de onda del sistema. Por tanto es posible hallar el
comportamiento en la regi\'on umbral ($Z_{core} \approx
Z_{core}^{(c)}$) por el m\'etodo perturbativo, entendiendo la
variaci\'on en el potencial de Coulomb con respecto a $Z_{core}$
como perturbaci\'on $\Delta{\hat{V}}$. Para ello se parte de la funci\'on de onda del
estado b\'asico $\psi^{(c)}$ y el Hamiltoniano en el umbral $\hat{H}(Z_{core}^{(c)})=\hat{H}^{(c)}$, y se procede como
sigue:

\begin{eqnarray}\label{comp_lineal}
\nonumber E &\approx& \langle \psi^{(c)}|\hat{H}^{(c)}|\psi^{(c)}\rangle + \langle \psi^{(c)}|\Delta{\hat{V}}|\psi^{(c)}\rangle\\
\nonumber &=& E(Z_{core}^{(c)}) + \langle \psi^{(c)}|-\frac{Z_{core}-Z_{core}^{(c)}}{r}|\psi^{(c)}\rangle \\
\nonumber &=& -\langle \psi^{(c)}| \frac{Z_{core}-Z_{core}^{(c)}}{r}|\psi^{(c)}\rangle \\
&=& c_{1} + c_{2} Z_{core}
\end{eqnarray}

\noindent N\'otese adem\'as que, el valor medio, bajo la funci\'on de onda en el umbral $\psi^{(c)}$, del Hamiltoniano evaluado en $Z^{(c)}$ es igual a la energ\'ia del electr\'on para $Z_{core}=Z_{core}^{(c)}$, o sea $\langle \psi^{(c)}|\hat{H}^{(c)}|\psi^{(c)}\rangle = E(Z_{core}^{(c)})$, y que, por definici\'on de $Z_{core}^{(c)}$, $E(Z_{core}^{(c)}) \equiv 0$.

Por tanto, el comportamiento de $E_{I}$ cercano al umbral de autoionizaci\'on es lineal con
$Z_{core}$. Este resultado y el hecho de que existe un radio finito
para la funci\'on de onda del \'ultimo electr\'on en el estado de
cero energ\'ia del umbral, indica que puede existir una transici\'on
de fase de primer orden en $Z_{core}^{(c)}$ para \'atomos con $N=2,
10, 18$ (en \cite{11} ya se hab\'ia se\~nalado esta particularidad
para los \'atomos con $N=2$).

\begin{figure}[htb]
  \begin{center}
  \includegraphics[width= .7\linewidth,angle=0]{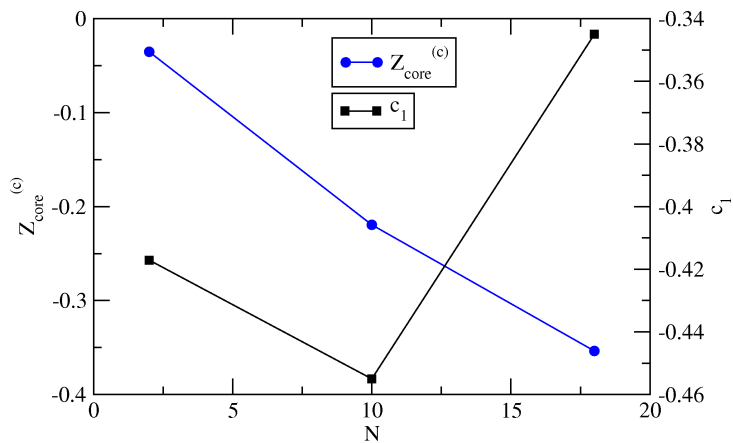}
  \end{center}
  \caption{a) Comportamiento de $Z_{core}^{(c)}$ (c\'irculos azules) y la pendiente $c_{1}$(cuadrados negros) con $N$.}\label{figura4}
\end{figure}

Los c\'alculos mostrados en la Tabla \ref{tabla_zc} fueron resultado
de un ajuste lineal de la data para $N=2, 10, 18$ para puntos
cercanos a $E_{I}=0$. En la Fig. \ref{figura4} se muestra la
relaci\'on entre $Z_{core}^{(c)}$ y $N$, y entre la pendiente $c_{1}$
(v\'ease la Ec. \ref{comp_lineal}) y $N$.

\noindent Los resultados de la Tabla \ref{tabla_zc} y la Fig.
\ref{figura4} pueden ser mejorados aumentando el tama\~no de la base
de $N$ part\'iculas (\ref{N_wavefunction}), o sea escogiendo un
$E_{corte}$ mayor que permita que la cantidad de configuraciones
crezca (v\'ease \ref{1p1h_selec_rules} y \ref{2p2h_selec_rules}).

\begin{figure}[h]
  \begin{center}
  \includegraphics[width= .7 \linewidth,angle=0]{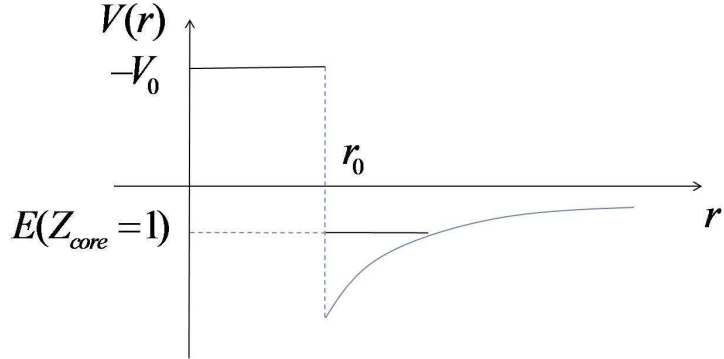}
  \end{center}
  \caption{Esquema que representa el potencial de una part\'icula caracter\'istico de los \'atomos con la estructura electr\'onica de los metales alcalinos.}\label{esquema_alcalinos}
\end{figure}

Ahora bien, analicemos el caso de los metales alcalinos. El
potencial de una part\'icula que describe correctamente este caso es tal que
para $Z_{core}<0$ no existe estado ligado (Fig. \ref{figura3} b) ),
En $Z_{core}=0$, $V(r)$ no debe admitir ning\'un nivel de energ\'ia,
el pozo cuadrado debe ser muy poco profundo o incluso pudiera
existir una barrera ($-V_{0}>0$) para $r<r_{0}$ (v\'ease la Fig.
\ref{esquema_alcalinos}).

El radio caracter\'istico de la interacci\'on de Coulomb es
inversamente proporcional a la carga nuclear (en este caso:
$r_{Coul}\sim 1/Z_{core}$), de manera que a cuando
$Z_{core}\rightarrow 0$ el radio $r_{Coul}$ crece infinitamente. O
sea para $Z_{core}$ suficientemente peque\~na $r_{Coul} \gg r_{0}$ y
se puede considerar entonces que $r_{0} \rightarrow 0$. Este
l\'imite permite hallar el comportamiento anal\'itico de $E_{I}$
cerca de $Z_{core}^{(c)}$ en el caso de los sistemas con la estructura electr\'onica de los metales alcalinos.

La funci\'on de onda radial para el l\'imite $r_{0}\rightarrow0$
resulta (v\'ease \ref{wavefunction_model}):

\begin{equation}
\begin{array}{cc}
R(r)= B \frac{W_{\beta,1/2}(\alpha r)}{r}, & r>r_{0}\rightarrow 0
\Rightarrow r>0
\end{array}
\end{equation}

Para cualquier problema de campo central la funci\'on de onda
evaluada en cero es estrictamente cero, de manera que:

\begin{equation}
W_{\beta(\varepsilon),1/2} (\alpha(\varepsilon) r_{0})=0
\end{equation}

Exigiendo que $\varepsilon \rightarrow 0$ (umbral de autoionizaci\'on) se
tiene que $\alpha(\varepsilon) \rightarrow 0$ (Ec. \ref{betaalpha}),
por tanto:

\begin{equation}
W_{\beta(\varepsilon),1/2} (0)=0
\end{equation}

La ecuaci\'on anterior s\'olo se cumple cuando $\beta(\varepsilon)$ es
entero. De la Ec. \ref{betaalpha}:

\begin{equation}
|\varepsilon|=\frac{1}{2}{\bigg (\frac{Z_{core}}{\beta} \bigg )}^{2}
\end{equation}

Ahora bien evaluando en esta expresi\'on el entero $\beta$ de menor
m\'odulo se obtiene la energ\'ia de mayor m\'odulo, i.e. la energ\'ia del
estado b\'asico o energ\'ia de enlace del electr\'on d\'ebilmente ligado. O
sea:

\begin{equation}
\nonumber |\varepsilon_{0}|=\frac{1}{2}{\bigg
(\frac{Z_{core}}{{|\beta|}_{min}} \bigg )}^{2}
\end{equation}

\begin{equation}\label{e_hidrogenoides}
\Rightarrow \varepsilon_{0}=-\frac{{Z_{core}}^{2}}{2}
\end{equation}

De manera que el comportamiento de la energ\'ia de ionizaci\'on cercano a
$Z_{core}^{(c)}$ es como el de los \'atomos hidrogenoides. Por tanto, $Z_{core}^{(c)}=0$ (\ref{e_hidrogenoides}). Hecho que
explica la inestabilidad de aniones at\'omicos para $N=3, 11, 19$ e
indica la zona universal en el comportamiento de $E_{I}$ vs.
$Z_{core}$, para $Z_{core} \approx 0$. En la Fig. \ref{figura3} esta universalidad no es
evidente porque los c\'alculos num\'ericos cercanos a
$Z_{core}^{(c)}$ (en este caso $Z_{core}^{(c)}=0$, v\'ease la Ec.
\ref{e_hidrogenoides}) son muy imprecisos. Esto \'ultimo es
consecuencia del tama\~no de la base de orbitales de una part\'icula
empleada; no se consideran funciones de onda de una part\'icula de
radio caracter\'istico suficientemente grande. Sin embargo, el
comportamiento anal\'itico hallado indica que el tipo de
transici\'on de fase presente en este caso ($N=3, 11, 19$) es
continua o de segundo orden (resultado obtenido en \cite{11} para
$N=3$).

%% file: conclusiones.tex
\chapter*{Conclusiones}
\addcontentsline{toc}{chapter}{Conclusiones}
\markright{CONCLUSIONES}

En la tesis presentada se observa ante todo que la estabilidad de aniones at\'omicos simplemente cargados es una propiedad dependiente del n\'umero de electrones de una forma cuasiperi\'odica, donde el per\'iodo var\'ia seg\'un el n\'umero de electrones que llenan una capa electr\'onica (e.g. $2, 8, 8$, etc.). Los \'atomos con la estructura electr\'onica de los gases nobles, son siempre m\'as estables (i.e. tienen energ\'ia de ionizaci\'on menor para todo $Z$) que los que tienen la estructura electr\'onica de los metales alcalinos, y s\'olo existen de forma estable a\-nio\-nes at\'omicos simplemente cargados ($Z=N-1$), en fase gaseosa, en el caso de los gases nobles. A pesar de que resulta evidente que $Z_{core}^{(c)}$ y $c_{1}$ deben ser funciones de $N$, dado que es el \'unico par\'ametro libre de los sistemas at\'omicos en este estudio, no se pudo obtener una dependencia de esta magnitud que fuera \'util para escalar los valores de $E_{I}$ y $Z_{core}$ en la Fig. \ref{figura3}. Para encontrar esta dependencia ser\'a necesario aumentar la cantidad de casos estudiados, o sea realizar c\'alculos num\'ericos de la energ\'ia de ionizaci\'on para $N=36, 54, 86$. 

El comportamiento de la energ\'ia de ionizaci\'on para el dominio ani\'onico es una ca\-rac\-te\-r\'is\-ti\-ca comparativamente diferente para estos dos grupos de \'atomos. Es notable, por ejemplo, que la aproximaci\'on de la energ\'ia de ionizaci\'on al umbral de la carga nuclear es mucho m\'as abrupta para los gases nobles, con una escala de variaci\'on de $\sim 0.9$ $u.a.$, mientras que en el caso de los metales alcalinos la variaci\'on es de $\sim 0.2$ $u.a.$. La dependencia anal\'itica, obtenida a partir del modelo de una part\'icula, de la energ\'ia de ionizaci\'on con la carga del cori\'on, en las proximidades de la carga cr\'itica del cori\'on, es lineal en el caso de los gases nobles, y cuadr\'atica (similar a la de los \'atomos hidrogenoides) en el caso de los metales alcalinos, lo que se ajusta a los resultados obtenidos mediante el c\'alculo num\'erico de CISD (Fig. \ref{figura3}) y se\~nala lo apropiado que es el modelo propuesto para la descripci\'on de los procesos de autoionizaci\'on en un \'atomo con $N$ electrones. 

La zona universal develada por el modelo de una part\'icula, en cada caso, es la zona pr\'oxima al umbral de autoionizaci\'on, o sea cercano a $Z_{core}^{(c)}$. En el caso de los metales alcalinos se tiene que, seg\'un el modelo de una part\'icula, $Z_{core}^{(c)}=0$. Aunque los c\'alculos num\'ericos no muestran esta particularidad correctamente, para algunos casos, es posible que esto se deba a la implementaci\'on computacional de los m\'etodos de la Qu\'imica Cu\'antica empleados y no a una propiedad f\'isica de estos sistemas.

%% file: recomendaciones.tex
\chapter*{Recomendaciones}
\addcontentsline{toc}{chapter}{Recomendaciones}
\markright{RECOMENDACIONES}

Se recomienda, sobre todo, mejorar los c\'alculos de la energ\'ia de ionizaci\'on de todos los sistemas considerados aumentando el tama\~no de la base de una part\'icula e incluyendo m\'as efectos de la correlaci\'on electr\'onica, i.e. elevando la dimensi\'on de la base de N part\'iculas (v\'eanse los ep\'igrafes \ref{corr}, \ref{CI} y \ref{comp}; y la Fig. \ref{cubic}), con el inconveniente del crecimiento del costo computacional. 

Se deber\'ian obtener los escalamientos adecuados de $E_{I}$ y $Z_{core}$ con $N$, de manera que las curvas de $E_{I}$ vs. $Z_{core}$ se concentren y la universalidad sea evidente. 

Adem\'as, ser\'ia \'util ampliar la cantidad de casos estudiados, e.g. incluir $N=36,54,86$ y $N=37,55,87$ en cada uno de los grupos comparados, y analizar los \'atomos con estructura electr\'onica similar a otros de los Grupos de la Tabla Peri\'odica.

%% file: appendixa.tex
\chapter{Elementos de matriz expl\'icitos en el m\'etodo de CISD}
\markright{AP\'ENDICE A}
\label{appA}

Para hallar expl\'icitamente los elementos de matriz que aparecen en el problema de autovalores y autovectores $\HH\textbf{C}=E \textbf{C}$ (\ref{autoval}), primeramente se utiliza la representaci\'on de los n\'umeros de ocupaci\'on para la funci\'on de onda $\Phi_{0}$ de HF (\ref{N_wavefunction}). Como por definici\'on la funci\'on de onda de HF es aquella en la que los electrones ocupan los orbitales de menor e\-ner\-g\'ia, esta se representa como $|HF\rangle = |1, 1, \ldots , 1, 0, 0 \ldots 0\rangle$ donde los n\'umeros $1$ se\~nalan los estados ocupados y los $0$, los desocupados o virtuales; en este caso, los primeros $N$ orbitales est\'an ocupados y el resto desocupados. 

A partir de la funci\'on $|HF\rangle$ se generan las configuraciones simples $\Phi_{\lambda}^{\sigma}$ y dobles $\Phi_{\lambda \mu}^{\sigma \tau}$ (que denotaremos en este ap\'endice por $|\lambda, \sigma \rangle$ y $|\lambda, \mu, \sigma, \tau \rangle$) aplicando los operadores de creaci\'on $\hat{e}^{\dagger}_{\sigma}$ y aniquilaci\'on $\hat{e}_{\lambda}$ de electrones en los estados $\sigma$ y $\lambda$, respectivamente. De modo que, si $\varepsilon_{f}$ es la energ\'ia de Fermi del sistema, en $|\lambda, \sigma \rangle = \hat{e}^{\dagger}_{\sigma} \hat{e}_{\lambda}|HF \rangle$ se ha eliminado un electr\'on en el estado $\lambda$, con $\varepsilon_{\lambda}<\varepsilon_{f}$, y se ha creado un electr\'on en el estado $\sigma$, con $\varepsilon_{\sigma}>\varepsilon_{f}$, y en $|\lambda, \mu, \sigma, \tau \rangle = \hat{e}^{\dagger}_{\tau} \hat{e}^{\dagger}_{\sigma} \hat{e}_{\mu} \hat{e}_{\lambda}|HF \rangle$ se han eliminado dos electrones en los estados $\lambda$ y $\mu$ ($\mu<\lambda$), con $\varepsilon_{\mu}, \varepsilon_{\lambda}<\varepsilon_{f}$, y se han creado dos electrones en los estados $\sigma$ y $\tau$ ($\tau<\sigma$), con $\varepsilon_{\sigma}, \varepsilon_{\tau}>\varepsilon_{f}$ (v\'ease Ep. \ref{CI}). 

Una vez escrito el Hamiltoniano en funci\'on de los operadores de creaci\'on y aniquilaci\'on, 

$$\hat{H}=\sum_{i,j}^{N}{\langle i |\hat{T}| j \rangle} \hat{e}^{\dagger}_{i} \hat{e}_{j} +\frac{1}{2}\sum_{i,j,k,l}^{N}{\langle i j |\hat{T}| k l \rangle} \hat{e}^{\dagger}_{i} \hat{e}^{\dagger}_{j} \hat{e}_{l} \hat{e}_{k}$$

\noindent las magnitudes:

$$D_{HF,\sigma\tau\mu\lambda}= \langle HF |\hat{H} \hat{e}^{\dagger}_{\tau} \hat{e}^{\dagger}_{\sigma} \hat{e}_{\mu} \hat{e}_{\lambda}|HF \rangle$$

$$B_{\sigma'\lambda',\sigma\tau\mu\lambda}=\langle HF | \hat{e}_{\lambda'} \hat{e}^{\dagger}_{\sigma'}\hat{H} \hat{e}^{\dagger}_{\tau} \hat{e}^{\dagger}_{\sigma} \hat{e}_{\mu} \hat{e}_{\lambda}|HF \rangle$$

$$A_{\sigma'\lambda',\sigma\lambda}=\langle HF | \hat{e}_{\lambda'} \hat{e}^{\dagger}_{\sigma'}\hat{H} \hat{e}^{\dagger}_{\sigma} \hat{e}_{\lambda}|HF \rangle$$

$$C_{\sigma'\tau'\mu'\lambda',\sigma\tau\mu\lambda}=\langle HF | \hat{e}_{\lambda'} \hat{e}_{\mu'} \hat{e}^{\dagger}_{\sigma'} \hat{e}^{\dagger}_{\tau'}\hat{H} \hat{e}^{\dagger}_{\tau} \hat{e}^{\dagger}_{\sigma} \hat{e}_{\mu} \hat{e}_{\lambda}|HF \rangle$$

\noindent se calculan usando el Teorema de Wick.

De manera que para la matriz de Tamm-Dankoff se tiene la expresi\'on:

\begin{eqnarray}
A_{\sigma'\mu',\sigma\mu}&=&\left(E_{HF}+\varepsilon^{(e)}_{\sigma}-
 \varepsilon^{(e)}_{\mu}\right)\delta_{\sigma\sigma'}\delta_{\mu\mu'}
 \nonumber\\
  &+& \langle\sigma',\mu|1/r_{ee}|\widetilde{\mu',\sigma}\rangle,
\end{eqnarray}

\noindent donde los elementos antisimetrizados de Coulomb se definen como:

\begin{equation}
\langle\sigma',\mu|1/r_{ee}|\widetilde{\mu',\sigma}\rangle=
\langle\sigma',\mu|1/r_{ee}|\mu',\sigma\rangle-
\langle\sigma',\mu|1/r_{ee}|\sigma,\mu'\rangle
\end{equation}

Los elementos de matriz de Coulomb $\langle\sigma',\mu|1/r_{ee}|\mu',\sigma\rangle$
son calculados en t\'erminos de los elementos de matriz sobre estados hidrogenoides (\ref{hidrogenoid}) usando una desarrollo similar a (\ref{base}).

Finalmente, las matrices $D$, $B$ y $C$ se escriben expl\'icitamente como:

\begin{equation}
D_{HF,\sigma\tau\mu\lambda}= \langle\mu,\lambda|1/r_{ee}|
 \widetilde{\tau,\sigma}\rangle
\end{equation}

\begin{eqnarray}
B_{\sigma'\mu',\sigma\tau\mu\lambda}&=&
 \langle\mu,\lambda|1/r_{ee}|\widetilde{\mu',\tau}\rangle
 \delta_{\sigma\sigma'} \nonumber\\
 &+&\langle\mu,\lambda|1/r_{ee}|\widetilde{\sigma,\mu'}\rangle
 \delta_{\tau\sigma'} \nonumber\\
 &+&\langle\sigma',\lambda|1/r_{ee}|\widetilde{\tau,\sigma}\rangle
 \delta_{\mu\mu'} \nonumber\\
 &+&\langle\sigma',\mu|1/r_{ee}|\widetilde{\sigma,\tau}\rangle
 \delta_{\lambda\mu'}
\end{eqnarray}

\begin{eqnarray}
&&C_{\sigma'\tau'\mu'\lambda',\sigma\tau\mu\lambda}=\nonumber\\
 &&\left ( E_{HF}+\varepsilon^{(e)}_{\sigma}+\varepsilon^{(e)}_{\tau}
 -\varepsilon^{(e)}_{\mu}-\varepsilon^{(e)}_{\lambda}\right )
 \delta_{\sigma\sigma'}\delta_{\tau\tau'}\delta_{\mu\mu'}
 \delta_{\lambda\lambda'}\nonumber\\
 &&+\langle\mu,\lambda|1/r_{ee}|\widetilde{\mu',\lambda'}\rangle
 \delta_{\sigma\sigma'}\delta_{\tau\tau'}
 +\langle\sigma',\mu|1/r_{ee}|\widetilde{\mu',\sigma}\rangle
 \delta_{\lambda\lambda'}\delta_{\tau\tau'}\nonumber\\
 &&+\langle\tau',\lambda|1/r_{ee}|\widetilde{\lambda',\tau}\rangle
 \delta_{\sigma\sigma'}\delta_{\mu\mu'}
 +\langle\tau',\mu|1/r_{ee}|\widetilde{\tau,\lambda'}\rangle
 \delta_{\sigma\sigma'}\delta_{\lambda\mu'}\nonumber\\
 &&+\langle\tau',\lambda|1/r_{ee}|\widetilde{\tau,\mu'}\rangle
 \delta_{\sigma\sigma'}\delta_{\mu\lambda'}
 +\langle\tau',\mu|1/r_{ee}|\widetilde{\mu',\tau}\rangle
 \delta_{\sigma\sigma'}\delta_{\lambda\lambda'}\nonumber\\
 &&+\langle\tau',\lambda|1/r_{ee}|\widetilde{\sigma,\lambda'}\rangle
 \delta_{\mu\mu'}\delta_{\tau\sigma'}
 +\langle\sigma',\lambda|1/r_{ee}|\widetilde{\tau,\lambda'}\rangle
 \delta_{\mu\mu'}\delta_{\sigma\tau'}\nonumber\\
 &&+\langle\sigma',\lambda|1/r_{ee}|\widetilde{\lambda',\sigma}\rangle
 \delta_{\mu\mu'}\delta_{\tau\tau'}
 +\langle\tau',\sigma'|1/r_{ee}|\widetilde{\tau,\sigma}\rangle
 \delta_{\mu\mu'}\delta_{\lambda\lambda'}\nonumber\\
 &&+\langle\tau',\mu|1/r_{ee}|\widetilde{\lambda',\sigma}\rangle
 \delta_{\lambda\mu'}\delta_{\tau\sigma'}
 +\langle\sigma',\mu|1/r_{ee}|\widetilde{\lambda',\tau}\rangle
 \delta_{\lambda\mu'}\delta_{\sigma\tau'}\nonumber\\
 &&+\langle\sigma',\mu|1/r_{ee}|\widetilde{\sigma,\lambda'}\rangle
 \delta_{\lambda\mu'}\delta_{\tau\tau'}
 +\langle\tau',\lambda|1/r_{ee}|\widetilde{\mu',\sigma}\rangle
 \delta_{\mu\lambda'}\delta_{\tau\sigma'}\nonumber\\
 &&+\langle\sigma',\lambda|1/r_{ee}|\widetilde{\sigma,\mu'}\rangle
 \delta_{\mu\lambda'}\delta_{\tau\tau'}
 +\langle\sigma',\lambda|1/r_{ee}|\widetilde{\mu',\tau}\rangle
 \delta_{\mu,\lambda'}\delta_{\sigma\tau'}\nonumber\\
 &&+\langle\tau',\mu|1/r_{ee}|\widetilde{\sigma,\mu'}\rangle
 \delta_{\lambda\lambda'}\delta_{\tau\sigma'}
 +\langle\sigma',\mu|1/r_{ee}|\widetilde{\tau,\mu'}\rangle
 \delta_{\lambda\lambda'}\delta_{\sigma\tau'}
\end{eqnarray}

%% file: appendixc.tex
\chapter{Elementos de matriz de Coulomb (integrales bielectr\'onicas) entre estados hidrogenoideos}
\markright{AP\'ENDICE C}
\label{appC}

Los elementos de matriz de Coulomb, definidos en (\ref{coulomb}), que entran en las ecuaciones de HF (\ref{HF}), pueden obtenerse anal\'iticamente si se escogen los estados hidrogenoides (\ref{hidrogenoid}) como la base de una part\'icula $\{\varphi\}$. De manera que, calculando la integral bielectr\'onica sobre estos estados,

\begin{equation}
\langle a, b || c, d \rangle = \int\int \{\varphi_{a}^{*}(r_{1}) \varphi_{b}^{*}(r_{2})\frac{1}{|r_{1}-r_{2}|} \varphi_{c}(r_{1}) \varphi_{d}(r_{2}) d^{3}r_{1} d^{3}r_{2}\} \delta_{S_{z_{a}} S_{z_{c}}} \delta_{S_{z_{b}}S_{z_{d}}}
\end{equation}

\noindent se obtiene:

\begin{eqnarray}\label{coulomb_explicit}
&&\langle a,b||c,d\rangle=Z_A \delta(s_{za},s_{zc})\delta(s_{zb},s_{zd}) 
 \delta(m_a+m_b,m_c+m_d)\delta(l_a+l_b+l_c+l_d,par) \nonumber \\
&\times& (-1)^{m_b-m_c} \sqrt{(2 l_a+1)(2 l_b+1)(2 l_c+1)(2 l_d+1)}\nonumber\\
&\times& \sqrt{(n_a-l_a-1)!(n_a+l_a)!(n_b-l_b-1)!(n_b+l_b)!(n_c-l_c-1)!(n_c+l_c)!(n_d-l_d-1)!(n_d+l_d)!}\nonumber\\
&\times& \sum^{l_{max}}_{l=l_{min}} \left(\begin{array}{ccc} l_a & l & l_c \\
 -m_a & m_a-m_c & m_c\end{array}\right) \left(\begin{array}{ccc}
 l_a & l & l_c \\
 0 & 0   & 0\end{array}\right)\left(\begin{array}{ccc}
 l_b & l & l_d \\
 -m_b & m_b-m_d & m_d\end{array}\right)\left(\begin{array}{ccc}
 l_b & l & l_d \\
 0 & 0   & 0\end{array}\right)\nonumber\\
&\times& \sum_{k_a=0}^{n_a-l_a-1}\sum_{k_b=0}^{n_b-l_b-1}
 \sum_{k_c=0}^{n_c-l_c-1}\sum_{k_d=0}^{n_d-l_d-1}
 \frac{(-1)^{k_a+k_b+k_c+k_d}~2^{k_a+k_b+k_c+k_d+l_a+l_b+l_c+l_d+4}}
 {k_a!k_b!k_c!k_d!}\nonumber\\
&\times& \frac{1}{n_a^{k_a+l_a+2}(n_a-l_a-k_a-1)!(2 l_a+k_a+1)!}~
 \frac{1}{n_b^{k_b+l_b+2}(n_b-l_b-k_b-1)!(2 l_b+k_b+1)!}\nonumber\\
&\times&\frac{1}{n_c^{k_c+l_c+2}(n_c-l_c-k_c-1)!(2 l_c+k_c+1)!}~
 \frac{1}{n_d^{k_d+l_d+2}(n_d-l_d-k_d-1)!(2 l_d+k_d+1)!}\nonumber\\
&\times& \left[\frac{(l+l_b+l_d+k_b+k_d+2)!}
 {(1/n_b+1/n_d)^{l+l_b+l_d+k_b+k_d+3}} \left\{\frac{(l_a+l_c+k_a+k_c+1-l)!}
 {(1/n_a+1/n_c)^{l_a+l_c+k_a+k_c+2-l}}\right.\right.\nonumber\\
&-&\left.\sum_{s=0}^{l+l_b+l_d+k_b+k_d+2}
 \frac{(1/n_b+1/n_d)^s(l_a+l_c+k_a+k_c+1+s-l)!}
 {s!(1/n_a+1/n_b+1/n_c+1/n_d)^{l_a+l_c+k_a+k_c+2+s-l}} \right\}\nonumber\\
&+&\frac{(l+l_a+l_c+k_a+k_c+2)!}{(1/n_a+1/n_c)^{l+l_a+l_c+k_a+k_c+3}} 
 \left\{\frac{(l_b+l_d+k_b+k_d+1-l)!}{(1/n_b+1/n_d)^{l_b+l_d+k_b+k_d+2-l}}
 \right.\nonumber\\
&-&\left.\left.\sum_{s=0}^{l+l_a+l_c+k_a+k_c+2}
 \frac{(1/n_a+1/n_c)^s(l_b+l_d+k_b+k_d+1+s-l)!}
 {s!(1/n_a+1/n_b+1/n_c+1/n_d)^{l_b+l_d+k_b+k_d+2+s-l}}\right\}\right].
\end{eqnarray}

\noindent
Aqu\'i $\left(\begin{array}{ccc} l_a & l & l_c\\ -m_a  & m & m_c 
\end{array}\right)$ son los denominados simbolos 3J, distintos de cero s\'olo 
cuando 

\begin{equation}
-m_a+m+m_c=0
\end{equation}

\noindent $l_a$, $l$, $l_c$ est\'an relacionados entre s\'i a trav\'es de las siguientes desigualdades triangulares:

\begin{equation}
\Delta(l_a,l,l_c) : \left\{\begin{array}{c} 
 l_a+l\ge l_c\\ l_c+l_a\ge l\\l+l_c\ge l_a \end{array}\right.
\end{equation}

\noindent y la sumatoria por $l$ en (\ref{coulomb_explicit}) corre desde $l_{min}=Max[|l_a-l_c|,|l_b-l_d|,|m_a-m_c|]$ hasta $l_{max}=Min[l_a+l_c,l_b+l_d]$, con $l+l_a+l_c=par$.

De la expresi\'on para $\langle a,b||c,d\rangle$ puede verse que la dependencia con $Z$ es trivial:

\begin{equation}
\langle a,b||c,d\rangle_Z=Z \langle a,b||c,d\rangle_{Z=1}
\end{equation}

\noindent As\'i que basta calcular los elementos de matriz para el Hidr\'ogeno y multiplicarlos por $Z$ en caso de cualquier otro elemento qu\'imico. 

En la implementaci\'on se hizo una tabla con los elementos de matriz distintos de cero para las integrales bielectr\'onicas entre todos los estados considerados, desde el $1s$ al $8j$, para optimizar la ejecuci\'on de los algoritmos de HF y CISD.

%% file: appendixb.tex
\chapter{Algoritmo de Lanczos para diagonalizar matrices sim\'etricas}
\markright{AP\'ENDICE B}
\label{appB}

El algoritmo de Lanczos es un algoritmo iterativo que resulta de una adaptaci\'on del m\'etodo de potencias para hallar autovalores y autovectores de una matriz sim\'etrica. Es particularmente \'util para diagonalizar matrices dispersas (con muchos elementos nulos) de dimensi\'on grande.

A continuaci\'on veremos algo de teor\'ia sobre el problema sim\'etrico de autovalores que utilizaremos luego para la construcci\'on del algoritmo de Lanczos.

\subsection{Algoritmo de Rayleigh-Ritz}

El m\'etodo de Lanczos se basa a su vez en el algoritmo de Rayleigh-Ritz para hallar una aproximaci\'on al autovalor m\'inimo $\lambda_{1}[A]$ y al m\'aximo $\lambda_{n}[A]$ de una matriz $A$. 

Primeramente enunciemos dos teoremas:

\begin{theorem}\label{Rayleigh}(Teorema del Coeficiente de Rayleigh)
Sea $A$ una matriz sim\'etrica $n \times n$, con autovalores en orden ascendente $\lambda_{1}[A] \ge \ldots \ge \lambda_{n}[A]$ y autovectores $\textbf{q}_{1},\ldots, \textbf{q}_{n}$, donde $A\textbf{q}_{i}=\lambda_{i}\textbf{q}_{i}$ $\forall{1<i<n}$. Sea $\textbf{x}$ un vector arbitrario tal que $\textbf{x} \in \R^{n}$ y sea el coeficiente de Rayleigh $\rho(\textbf{x};A)=\frac{\textbf{x}^{T}A\textbf{x}}{\textbf{x}^{T}\textbf{x}}$, entonces:

\begin{eqnarray}
&&\min_{x\in\R^{n}}{\rho(\textbf{x};A)}=\rho(\textbf{q}_{1};A)=\lambda_{1}[A]\\
&&\max_{x\in\R^{n}}{\rho(\textbf{x};A)}=\rho(\textbf{q}_{n};A)=\lambda_{n}[A] 
\end{eqnarray}

\end{theorem}

%\begin{proof}

\noindent \textbf{Demostraci\'on \ref{Rayleigh}:}

\noindent Sea el vector unitario $\textbf{u}=\frac{\textbf{x}}{\|\textbf{x}\|}$ tal que $\|\textbf{u}\|=1$, si se parte de la definici\'on del coeficiente de Rayleigh, y se tiene en cuenta que $\textbf{x}^{T} \cdotp \textbf{x}=\|\textbf{x}\|^{2}$, se puede obtener:

\begin{eqnarray}\label{expr1}
\rho(\textbf{x};A)&=&\frac{\textbf{x}^{T}A\textbf{x}}{\textbf{x}^{T}\textbf{x}}\nonumber \\
&=&\frac{\textbf{x}^{T}A\textbf{x}}{\|\textbf{x}\|^{2}}\nonumber \\
&=&\textbf{u}^{T}A\textbf{u}
\end{eqnarray}

\noindent Ahora bien, hallemos el m\'inimo de la funci\'on de varias variables $\rho(\textbf{x};A)$. Usando (\ref{expr1}) y considerando la restricci\'on $\textbf{u}^{T} \cdotp \textbf{u}=\|\textbf{u}\|^{2}=1$ se puede plantear el problema de extremos condicionados para el funcional:

\begin{equation}
\frak{L}(\textbf{u})=\textbf{u}^{T}A\textbf{u}-\lambda(\textbf{u}^{T} \cdotp \textbf{u}-1)
\end{equation}

\noindent Utilizando el convenio de Einstein para \'indices de sumaci\'on, se puede escribir:

\begin{equation}\label{einst}
\frak{L}(\textbf{u})=u_{i}a_{ij}u_{j}-\lambda(u_{i}u_{i}-1)
\end{equation}

\noindent donde $\textbf{u}=\lVert u_{i} \rVert_{n}$ y $A=\lVert a_{ij} \rVert_{n \times n}$.

\noindent De manera que el problema de extremos condicionados, haciendo un poco de \'algebra,

\begin{eqnarray}\label{einst}
\nonumber \frac{\partial{\frak{L}(\textbf{u})}}{\partial{u_{k}}}&=&u_{i}a_{ij}u_{j}-\lambda(u_{i}u_{i}-1)\\
\nonumber &=&a_{kj}u_{j} + u_{i}a_{ik} -2 \lambda u_{k}\\
\nonumber &=&a_{kj}u_{j} + a_{ik}u_{i} -2 \lambda u_{k}\\
\nonumber &=&a_{kj}u_{j} + a_{ki}u_{i} -2 \lambda u_{k} ; (a_{ki}=a_{ik})\\
\nonumber &=&a_{ij}u_{j} + a_{ij}u_{j} -2 \lambda u_{j} ; (i \geq j, k \geq i) \\
&=&2a_{ij}u_{j} -2 \lambda u_{j}
\end{eqnarray}

\noindent se reduce a:

\begin{equation}\label{einst}
2a_{ij}u_{j} -2 \lambda u_{j}=0
\end{equation}

\noindent \'o:

\begin{eqnarray}\label{autoval_simetric}
&&2 A \textbf{u} - 2 \lambda \textbf{u} = \textbf{0} \nonumber \\
&&\Rightarrow A \textbf{u} = \lambda \textbf{u}
\end{eqnarray}

\noindent Por tanto, los vectores $\textbf{u}$ que representan los puntos estacionarios de $\rho(\textbf{x};A)$ son los autovectores normalizados de $A$, o sea $\textbf{u}_{i}=\textbf{q}_{i} / \|\textbf{q}_{i}\|$, $\forall_{1<i<n}$. O sea que:

\begin{equation}\label{expr2}
\rho(\textbf{x};A)=\textbf{u}^{T}A\textbf{u}=\lambda \textbf{u}^{T}\textbf{u}=\lambda
\end{equation}

\noindent De manera que, los valores de $\rho(\textbf{x};A)$ en los puntos estacionarios $\textbf{u}$, son los autovalores $\lambda$ de $A$. Finalmente, se sigue que:

\begin{eqnarray}
&&\min_{x\in\R^{n}}{\rho(\textbf{x};A)}=\min_{\lambda \in \{ \lambda_{i} \}; 1<i<n}{\lambda} \equiv \lambda_{1}[A]\\
&&\max_{x\in\R^{n}}{\rho(\textbf{x};A)}=\max_{\lambda \in \{ \lambda_{i} \}; 1<i<n}{\lambda} \equiv \lambda_{n}[A] 
\end{eqnarray}

\noindent Que es lo que se quer\'ia demostrar. \EndProof

%\end{proof}

\begin{theorem}\label{Rayleigh-Ritz}(Teorema de Rayleigh-Ritz)
Sea $A$ una matriz sim\'etrica $n \times n$, con autovalores en orden ascendente $\lambda_{1}[A] \ge \ldots \ge \lambda_{n}[A]$. Sea $\{q_{1},\ldots,q_{m}\}$, con $q_{i}\in\R^{n} \forall_{1<i<m}$, una base ortonormal en el subespacio $\R^{m} (m<n) \subset \R^{n}$, de manera que cualquier vector $\textbf{x}$ tal que $\textbf{x} \in \R^{m}$ pueda ser expresado como combinaci\'on lineal de los vectores $\{q_{1},\ldots,q_{m}\}$:

\begin{equation}
\textbf{x}=z_{1}\textbf{q}_{1}+\ldots+z_{n}\textbf{q}_{n} \nonumber
\end{equation}

\noindent \'o:

\begin{equation}\label{comb_lineal}
 \left [ \begin{array}{c} x_{1} \\ \vdots \\ x_{n} \\ \end{array} \right ] =
 \left [\begin{array}{ccc}
     &  &  \\
    \textbf{q}_{1} & \ldots & \textbf{q}_{m} \\
     &  &  \\
  \end{array} \right ]
\left [ \begin{array}{c} z_{1} \\ \vdots \\ z_{m} \\ \end{array} \right ]
\end{equation}

\noindent Entonces la mejor aproximaci\'on para $\lambda_{1}[A]$ y $\lambda_{n}[A]$ se obtiene de diagonalizar $H=Q^{T}AQ$, y est\'a dada por $\lambda_{1}[H]$ y $\lambda_{n}[H]$, respectivamente.
\end{theorem}

%\begin{proof}

\noindent \textbf{Demostraci\'on \ref{Rayleigh-Ritz}:}

\noindent Si se emplea la expresi\'on (\ref{comb_lineal}) y se tiene en cuenta que $Q$ es ortogonal ($Q^{T}Q=I$, donde $I$ es la matriz identidad) se puede expresar el coeficiente de Rayleigh como:

\begin{eqnarray}
\rho(\textbf{x};A)&=&\frac{\textbf{x}^{T}A\textbf{x}}{\textbf{x}^{T}\textbf{x}} \nonumber \\
&=&\frac{\textbf{z}^{T}Q^{T}AQ\textbf{z}}{\textbf{z}^{T}Q^{T}Q\textbf{z}} \nonumber \\
&=&\frac{\textbf{z}^{T}H\textbf{z}}{\textbf{z}^{T}\textbf{z}}
\end{eqnarray}

\noindent De manera que $\rho(\textbf{x};A)=\rho(\textbf{z};H)$; por tanto, el m\'inimo (m\'aximo) del coeficiente de Rayleigh para $A$ ser\'a tambi\'en el m\'inimo (m\'aximo) del coeficiente de Rayleigh para $H$, que es justamente $\lambda_{1}[H]$ ($\lambda_{n}[H]$), seg\'un el Teorema \ref{Rayleigh}. Finalmente, se tiene que $\lambda_{1}[A]=\lambda_{1}[H]$ y $\lambda_{n}[A]=\lambda_{n}[H]$. Queda demostrado. \EndProof 

\vspace{0.2in}

%\end{proof}

A partir de este \'ultimo teorema se construye el algoritmo de Rayleigh-Ritz para hallar los autovalores m\'aximo y m\'inimo de una matriz sim\'etrica, adem\'as de los correspondientes autovectores. La idea fundamental es construir una matriz $H=Q^{T}AQ$ que sea mucho m\'as f\'acil de diagonalizar que $A$ y cuyos autovalores m\'aximo y m\'inimo coincidan. A continuaci\'on se detallan los pasos del algoritmo:

\begin{enumerate}
\item Se propone un conjunto de $m$ vectores arbitrarios $S=[s_{1},\ldots,s_{m}]$ tales que $s_{j} \in \R^{n}$ y se aplica la ortonormalizaci\'on de Gram-Schmidt para obtener un conjunto de vectores ortonormales $Q=[q_{1},\ldots,q_{m}]$.
\item Se construye la matriz $H=Q^{T}AQ$.
\item Se diagonaliza $H$ y se obtienen $\lambda_{1}[H],\ldots,\lambda_{m}[H]$, tales que $H\textbf{g}_{k}=\lambda_{k}[H]\textbf{g}{k}$, $k=1,\ldots,m$.
\item La mejor aproximaci\'on para los autovalores $\lambda_{1}[A]$ y $\lambda_{n}[A]$ est\'a dada por $\lambda_{1}[H]$ y $\lambda_{n}[H]$, con los correspondientes autovectores $y_{k}=Qg_{k}$.
\end{enumerate}

\subsection{Algoritmo de Lanczos}

El algoritmo de Lanczos usa impl\'icitamente el algoritmo de Rayleigh-Ritz, o sea propone una forma iterativa de hallar la matriz de vectores ortogonales $Q$ tales que $H$ sea tridiagonal. Existen algoritmos sencillos para diagonalizar matrices tridiagonales sim\'etricas de manera que obtener una matriz $H$ en esta forma es muy ventajoso computacionalmente.

A continuaci\'on veremos un teorema que permitir\'a entender los pasos del algoritmo de Lanczos:

%\begin{definition}(Subespacio de Krylov)
%Se define el subespacio de Krylov de orden $m$, denominado $S_{m}$, como el espacio generado por los vectores $\textbf{f}, %A\textbf{f}, A^{2}\textbf{f},\ldots, A^{m-1}\textbf{f}$, que conforman la matriz de Krylov de orden $m$ %$K_{m}(\textbf{f})=[\textbf{f} A \textbf{f} \ldots A^{m-1}\textbf{f}]$, donde $\textbf{f}$ es un vector arbitrario tal que %$\textbf{f}\in\R^{n}$.
%\end{definition}

%\begin{theorem}
%Sea $Q_{m}$ la base ortonormal obtenida por el m\'etodo de Gram-Schmidt a partir de los vectores de $K_{m}(f)$, entonces %$T_{m}=Q_{m}^{T}AQ$ es una matriz tridiagonal.
%\end{theorem}

\begin{theorem}\label{lanczos_matrix}
Sea $A$ una matriz sim\'etrica $n \times n$. Sea $Q_{m}=[\textbf{q}_{1}, \ldots, \textbf{q}_{m}]$ la base ortonormal obtenida por el m\'etodo de Gram-Schmidt: 

\begin{eqnarray}\label{Gram-Schmidt}
\textbf{q}_{1}&=&\textbf{s}_{1}/\|\textbf{s}_{1}\| \nonumber \\
\textbf{q'}_{i}&=&\textbf{s}_{i}-\sum_{k=1}^{i-1}{\textbf{q}_{k}(\textbf{q}_{k} \cdotp \textbf{s}_{i})} \\
\textbf{q}_{i}&=&\textbf{q'}_{i}/\|\textbf{q'}_{i}\| \hspace{0.1in} ; \hspace{0.1in}  \forall_{2<i<m} \nonumber
\end{eqnarray}

\noindent donde $\textbf{s}_{1}$ es un vector arbitrario tal que $\textbf{s}_{1} \in \R^{n}$ y $\textbf{s}_{i}=A\textbf{q}_{i-1}$ $\forall_{2<i<m}$. Entonces $T_{m}=Q_{m}^{T}AQ$ es una matriz tridiagonal.
\end{theorem}

\noindent \textbf{Demostraci\'on \ref{lanczos_matrix}:}

\noindent Veamos algunos detalles de la ortonormalizaci\'on de Gram-Schmidt. Primeramente, por la construcci\'on del m\'etodo se tiene que $\textbf{q}_{i} \cdotp \textbf{q}_{j}=0$ $\forall$, $i \ne j$. Tomando $j > i$ y utilizando (\ref{Gram-Schmidt}) se puede obtener:

\begin{eqnarray}\label{expr3}
\nonumber \textbf{q}_{i} \cdotp \textbf{q}_{j} &=& \frac{\textbf{s}_{i} \cdotp \textbf{q}_{j} -\sum_{k=1}^{i-1}{(\textbf{q}_{k} \cdotp \textbf{q}_{j})(\textbf{q}_{k} \cdotp \textbf{s}_{i})}}{\|\textbf{q'}_{i}\|} \equiv 0 \\
&=&\frac{\textbf{s}_{i} \cdotp \textbf{q}_{j}}{\|\textbf{q'}_{i}\|} = 0
\end{eqnarray}

\noindent N\'otese que, como por construcci\'on $\textbf{q}_{k} \cdotp \textbf{q}_{j}=0$ $\forall$ $k \ne j$, y como $1 < k < i-1$ y $j > i$, el segundo t\'ermino en (\ref{expr3}) es exactamente nulo y, por tanto, $\textbf{s}_{i} \cdotp \textbf{q}_{j}=0$.

\noindent Ahora bien, en el caso concreto del Teorema \ref{lanczos_matrix} se tiene que $\textbf{s}_{j+1}=A\textbf{q}_{j}$, de manera que, seg\'un (\ref{expr3}), si $i>j+1$ se cumple que $\textbf{q}_{i} \cdotp (A\textbf{q}_{j}) \equiv \textbf{q}_{i}^{T}(A\textbf{q}_{j}) =0$. Adem\'as, como $A$ es sim\'etrica ($A=A^{T}$) se tiene que $\textbf{q}_{j}^{T}(A\textbf{q}_{i})=(\textbf{q}_{i}^{T}(A^{T}\textbf{q}_{j}))^{T}=\textbf{q}_{i}^{T}(A^{T}\textbf{q}_{j})=\textbf{q}_{i}^{T}(A\textbf{q}_{j})=0$; por tanto, $\textbf{q}_{i}^{T}(A\textbf{q}_{j}) =0$ para $i > j+1$ y para $i < j-1$. Finalmente, se obtiene que los elementos de la matriz $T_{m}=Q_{m}^{T}AQ_{m}$, que expl\'icitamente se escriben como $\textbf{q}_{i}^{T}(A\textbf{q}_{j}$, son nulos para $i > j+1$ y para $i < j-1$, donde $i$ es el \'indice de las filas y $j$ el de las columnas. De manera que la matriz $T_{m}$ tiene forma tridiagonal:

\begin{equation}\label{comb_lineal}
 T_{m}=\left [\begin{array}{ccccc}
    \alpha_{1} & \beta_{1} & & & \\
    \beta_{1} & \alpha_{2} & \beta_{2} & &\\
     & \ddots & \ddots & \ddots &\\
     &  & \beta_{m-2} & \alpha_{m-2} & \beta_{m-1}\\
     &  &  & \beta_{m-1} & \alpha_{m}\\
  \end{array} \right ]
\end{equation}

\noindent donde:

$$\left \{ \begin{array}{cccc}
\alpha_{i} & = & \textbf{q}_{i}^{T}(A\textbf{q}_{i}) & i=1, \ldots, m \nonumber \\
\beta_{i} & = & \textbf{q}_{i+1}^{T}(A\textbf{q}_{i}) & i=1, \ldots, m-1  
\end{array} \right . $$

\noindent Queda demostrado. \EndProof

Dado que la matriz $T_{m}$ es tridiagonal (Teorema \ref{lanczos_matrix}), i.e. $\textbf{q}_{i}^{T}(A\textbf{q}_{j}) =0$ para $i > j+1$ y para $i < j-1$, resulta que $A\textbf{q}_{j}$ es una combinaci\'on lineal de los vectores $\textbf{q}_{j-1}$, $\textbf{q}_{j}$ y $\textbf{q}_{j+1}$, o sea:

\begin{equation}\label{trid_expr}
A\textbf{q}_{i}=\beta_{i-1}\textbf{q}_{i-1}+\alpha_{i}\textbf{q}_{i}+\beta_{i}\textbf{q}_{i+1} \hspace{0.2in} (2 \le i \le m-1)
\end{equation}

\noindent Si definimos $\textbf{q}_{0}=0$ la ecuaci\'on anterior tambi\'en es v\'alida para $i = 1$.

Ahora bien, si $r_{i}=\beta_{i}\textbf{q}_{i+1}$ ($r_{i}$ es una componente de $A\textbf{q}_{i}$ ortogonal a $\textbf{q}_{j}$ para $j \le i$), entonces de (\ref{trid_expr}):

\begin{equation}\label{algorithm}
r_{i}=A\textbf{q}_{i}-\beta_{i-1}\textbf{q}_{i-1}-\alpha_{i}\textbf{q}_{i} \hspace{0.2in} (2 \le i \le m-1)
\end{equation}

Empleando la expresi\'on (\ref{algorithm}) se puede construir el Algoritmo de Lanczos (\ref{lanczos}).

\begin{algorithm}[htbp]\label{lanczos}\caption{Algoritmo de Lanczos}

\begin{algorithmic}

\STATE $\textbf{r}_{0}=Random(\textbf{x} ; \textbf{x} \in \R^{n})$ 

\STATE $\beta_{0}=\|r_{0}\|$

\STATE $\textbf{q}_{0}=\textbf{0}$

\STATE $i=0$

\WHILE{($|(\lambda_{1}^{a} - \lambda_{1}^{a-1})/\lambda_{1}^{a-1}| \geq \Delta$ \textbf{or} $|(\lambda_{n}^{a} - \lambda_{n}^{a-1})/\lambda_{n}^{a-1}| \geq \Delta)$}

	\STATE $i \gets i+1$

	\STATE $\textbf{q}_{i} \gets r_{i-1}/\beta_{i-1}$  

	\STATE $\textbf{r}_{i} \gets A\textbf{q}_{i}-\beta_{i-1}\textbf{q}_{i-1}$

	\STATE $\alpha_{i} \gets \textbf{q}_{i}^{T}\textbf{r}_{i}$  
		
	\STATE $r_{i} \gets r_{i} - \alpha_{i}\textbf{q}_{i}$

	\STATE $\beta_{i}=\|r_{i}\|$

	\STATE Construir la matriz $T_{i+1}$ a partir de $\{\alpha_{i}\}$ y $\{\beta_{i}\}$.

	\STATE Diagonalizar $T_{i+1}$. Hallar autovalores ($\lambda_{1}[T_{i+1}]$ y $\lambda_{n}[T_{i+1}]$) y autovectores ($\textbf{v}_{1}[T_{i+1}]$ y $\textbf{v}_{n}[T_{i+1}]$).

	\STATE $\lambda_{1}^{i} \gets \lambda_{1}[T_{i+1}]$
	
	\STATE $\textbf{v}_{1}^{i} \gets \textbf{v}_{1}[T_{i+1}]$

	\STATE $\lambda_{n}^{i} \gets \lambda_{n}[T_{i+1}]$
	
	\STATE $\textbf{v}_{n}^{i} \gets \textbf{v}_{n}[T_{i+1}]$

\ENDWHILE

\STATE $\lambda_{1} \gets \lambda_{1}^{i}$

\STATE $\textbf{v}_{1} \gets \textbf{v}_{1}^{i}$

\STATE $\lambda_{n} \gets \lambda_{n}^{i}$

\STATE $\textbf{v}_{n} \gets \textbf{v}_{n}^{i}$

\end{algorithmic}
\end{algorithm}

Como puede verse, este algoritmo consiste en construir la matriz tridiagonal $T_{m}$ que ser\'a luego diagonalizada de manera exacta. Primeramente, se escoge un vector unitario aleatorio, luego se aplica la matriz sim\'etrica $A$ que se quiera diagonalizar y se crea un nuevo vector que ser\'a ortonormalizado con el inicial por el proceso de Gram-Schmidt. Aplicando la matriz al vector unitario ortonormal al inicial, se crea otro vector que deber\'a ortonormalizarse con respecto a los dos anteriores; as\'i sucesivamente, aplicando reiteradamente la matriz $A$, se construyen los dem\'as vectores mediante un esquema que reproduce la expresi\'on (\ref{trid_expr}). Este procedimiento permite obtener, en la iteraci\'on $m$-\'esima, los valores de $\{\alpha_{i}\}$ y $\{\beta_{i}\}$ y construir la matriz $T_{m}$. Una vez que esta matriz se diagonalice, se comprueba si existe convergencia para los autovalores (i.e. $|(\lambda_{1}^{a} - \lambda_{1}^{a-1})/\lambda_{1}^{a-1}| \leq \Delta$ y $|(\lambda_{n}^{a} - \lambda_{n}^{a-1})/\lambda_{n}^{a-1}| \leq \Delta$). Si han convergido, $\lambda_{1}[T_{m}]$ y $\lambda_{n}[T_{m}]$ son las mejores aproximaciones a los autovalores $\lambda_{1}[A]$ y $\lambda_{n}[A]$, respectivamente. Si no, se aumenta la dimensi\'on $m$ de la matriz, o sea se calculan m\'as vectores $\textbf{q}_{i}$ con $i > m$ que transformen $A$ en una matriz tridiagonal $T_{m}$, hasta que se obtenga convergencia.

\subsection{Estabilidad num\'erica del algoritmo de Lanczos}

Hemos probado que con aritm\'etica \textit{exacta}, los vectores $\{q_{i}\}$, construidos en el algoritmo de Lanczos, son una base ortonormal, y los autovalores y autovectores de $H$ son buenas aproximaciones a los autovalores y autovectores de la matriz original $A$. Sin embargo, en la pr\'actica, mientras el c\'alculo se ejecuta en aritm\'etica de punto flotante, donde la inexactitud es inevitable, la ortogonalidad se pierde r\'apidamente y en algunos casos el siguiente vector calculado es incluso linealmente dependiente de los anteriores. Como resultado, algunos de los autovalores de la matriz tridiagonal resultante pueden no ser aproximaciones de los autovalores de la matriz inicial. Por tanto, el algoritmo de Lanczos no es muy estable.

Los usuarios de este algoritmo deben ser capaces de encontrar y eliminar los autovalores espurios. Para ello las implementaciones pr\'acticas del algoritmo de Lanczos se dedican a:

\begin{itemize}

\item Evitar la p\'erdida de la ortogonalidad.

\item Reortogonalizar despu\'es de generada la base.

\item Identificar todos los autovalores espurios y eliminarlos. 

\end{itemize}

Como conservar la ortogonalidad de un n\'umero grande vectores es intratable, se suele mantener la cantidad de vectores acotada y si no se tiene convergencia de los autovalores para el valor m\'aximo iteraciones entonces existe la posibilidad de reiniciar el algoritmo y escoger un vector inicial diferente. Una de las variantes del algoritmo de Lanczos es el reinicio impl\'icito (\textit{implicitly restarted Lanczos algorithm}), que consiste en crear el vector inicial usando la informaci\'on de la matriz tridiagonal de la iteraci\'on anterior.%; una implementaci\'on de esta variante es la empleada para diagonalizar matrices en el presente trabajo.